\documentclass[
  a4paper,
  superscriptaddress,
  twocolumn,
  aps,
  prl,
  10pt,
  preprintnumbers,
  floatfix,
  secnumarabic,
  amssymb
]
{revtex4-2}

\usepackage{changes}
\usepackage{dcolumn}   
\usepackage{bm}        
\usepackage{amssymb}   
\usepackage{amsmath,stmaryrd}
\usepackage{blkarray, multirow, graphicx, diagbox, color, xcolor, colortbl}
\usepackage{bbm, bbold}
\usepackage[acronym]{glossaries}
\usepackage{hyphenat}
\usepackage{ifthen}
\usepackage{float}

\usepackage{xkeyval}
\usepackage{moreverb}
\usepackage{rotating}
\usepackage{wrapfig}
\usepackage{slashbox}
\usepackage{xspace}
\usepackage{nicefrac}

\usepackage{siunitx}
\usepackage{physics}
\usepackage{booktabs}
\usepackage{braket}
\usepackage{float}
\usepackage[inline]{enumitem}
\usepackage{tabto}
\usepackage{listings}
\usepackage{xstring}
\usepackage{lipsum}
\usepackage{scalerel}

\def\ReplaceStr#1{%
	\IfSubStr{#1}{p}{%
		\StrSubstitute{#1}{p}{.}}{#1}}

\usepackage[caption=false]{subfig}
\usepackage{calc}

\definecolor{colorA}{rgb} {0.58,0,0.8275}
\definecolor{colorB}{rgb} {0.11,0.663,0.51}
\definecolor{colorC}{rgb} {0.3373,0.7059,0.9137}
\definecolor{colorD}{rgb} {0.902,0.6235,0}
\definecolor{colorE}{rgb} {0.9451,0.902,0.3255}
\definecolor{colorF}{rgb} {0.3373,0.3255,0.902}
\definecolor{colorG}{rgb} {0.9451,0.3255,0.3373}
\definecolor{cbColorA}{HTML} {2D4C8D}
\definecolor{cbColorB}{HTML} {EB821D}
\definecolor{cbColorC}{HTML} {C3342F}
\definecolor{cbColorD}{HTML} {24451B}
\definecolor{cbColorE}{HTML} {2C0041}
\definecolor{cbColorF}{HTML} {D9D9D9}

\definecolor{stBlue}{HTML} {5566AA}
\definecolor{stGreen}{HTML} {117733}
\definecolor{stCyan}{HTML} {33BBEE}
\definecolor{stTeal}{HTML} {009988}
\definecolor{stOrange}{HTML} {EE7733}
\definecolor{stYellow}{HTML} {F7F056}
\definecolor{stRed}{HTML} {CC3311}
\definecolor{stMagenta}{HTML} {EE3377}
\definecolor{stGrey}{HTML} {BBBBBB}

\newboolean{buildCurrentBlockInline}
\setboolean{buildCurrentBlockInline}{false}
\newboolean{rebuildCurrentBlock}
\setboolean{rebuildCurrentBlock}{false}

\newboolean{rebuildData}
\setboolean{rebuildData}{false}

\newboolean{removeData}
\setboolean{removeData}{false}

\graphicspath{{figures/}}

\usepackage[normalem]{ulem}
\usepackage{xr-hyper}
\usepackage[colorlinks,bookmarks=false,citecolor=blue,linkcolor=black,urlcolor=blue]{hyperref}
\usepackage{orcidlink}
\newacronym[shortplural={MPS}]{MPS}{MPS}{matrix\hyp product state}
\newacronym{MPSs}{MPS}{matrix\hyp product states}
\newacronym[shortplural={PP-MPS}]{PP-MPS}{PP-MPS}{projected purified matrix\hyp product state}
\newacronym{PP-MPSs}{PP-MPS}{projected purified matrix\hyp product states}
\newacronym{MPO}{MPO}{matrix-product operator}
\newacronym{SVD}{SVD}{singular-value decomposition}
\newacronym{QCS}{QCS}{quantum-computer simulator}
\newacronym{FSM}{FSM}{finite-state machine}
\newacronym{ACA}{ACA}{adaptive cross approximation}
\newacronym{1D}{1D}{one\hyp dimensional}
\newacronym{QC}{QC}{quantum computer}
\newacronym{CDW}{CDW}{charge\hyp density wave}
\newacronym{BOW}{BOW}{bond\hyp order wave}
\newacronym{SDW}{SDW}{spin\hyp density wave}
\newacronym{ARPES}{ARPES}{angle-resolved photoemission spectroscopy}
\newacronym{OBC}{OBC}{open-boundary conditions}
\newacronym{PBC}{PBC}{periodic-boundary conditions}
\newacronym{TEBD}{TEBD}{time-evolving block-decimation}
\newacronym{TDVP}{TDVP}{time\hyp dependent variational principle}
\newacronym{iff}{iff}{if and only if}
\newacronym{DFT}{DFT}{density\hyp functional theory}
\newacronym{DMFT}{DMFT}{dynamical mean\hyp field theory}
\newacronym{DMRG}{DMRG}{density\hyp matrix renormalization group}
\newacronym{1DMRG}{1DMRG}{single-site density\hyp matrix renormalization group}
\newacronym{2DMRG}{2DMRG}{two-site density\hyp matrix renormalization group}
\newacronym{DMRG3S}{DMRG3S}{strictly single-site density\hyp matrix renormalization group}
\newacronym{iDMRG}{iDMRG}{inifinite\hyp size density\hyp matrix renormalization group}
\newacronym{tDMRG}{tDMRG}{time\hyp dependent density\hyp matrix renormalization group}
\newacronym{PP-DMRG}{PP-DMRG}{projected purified density\hyp matrix renormalization group}
\newacronym{QMC}{QMC}{quantum Monte Carlo}
\newacronym{AIM}{AIM}{Anderson impurity model}
\newacronym{SIAM}{SIAM}{single impurity Anderson model}
\newacronym{LDA}{LDA}{local\hyp density approximation}
\newacronym{LBNL}{LBNL}{Lawrence Berkeley National Laboratory}
\newacronym{ED}{ED}{exact diagonalization}
\newacronym{QPT}{QPT}{quantum phase transition}
\newacronym{QCP}{QCP}{quantum critical point}
\newacronym{ETH}{ETH}{eigenstate thermalization hypothesis}
\newacronym{EHM}{EHM}{extended Hubbard model}
\newacronym{AKLT}{AKLT}{Affleck\hyp Lieb\hyp Kennedy\hyp Tasaki}
\newglossaryentry{QR}{name={QR},description={QR decomposition}}
\newacronym{TNS}{TNS}{tensor\hyp network state}
\newacronym{TN}{TN}{tensor\hyp network}
\newacronym{SM}{SM}{supplemental material}
\newacronym{NOO}{NOO}{natural orbital occupation}
\newacronym{NO}{NO}{natural orbital}
\newacronym{LRO}{LRO}{long\hyp range order}
\newacronym{qLRO}{qLRO}{quasi\hyp long\hyp range order}
\newacronym{SC}{SC}{Superconductivity}
\newacronym{VBGS}{VBGS}{valence bond ground-state}
\newacronym{PEPS}{PEPS}{projected entangled pair\hyp states}
\newacronym{ALS}{ALS}{alternating least squares}
\newacronym{BdG}{BdG}{Bogoljubov de-Gennes}
\newacronym{TFIM}{TFI}{transverse field Ising model}
\newacronym{PP}{PP}{projected purification}
\newacronym{BEC}{BEC}{Bose\hyp Einstein condensate}
\newacronym{JWT}{JWT}{Jordan\hyp Wigner transformation}
\newacronym{NISQ}{NISQ}{noisy intermediate scale quantum}
\newacronym{NN}{NN}{nearest\hyp neighbor}
\newacronym{NNN}{NNN}{next\hyp nearest\hyp neighbor}
\newacronym{SPDM}{SPDM}{single\hyp particle density matrix} 
\newacronym{HCB}{HCB}{hardcore bosons}
\newacronym{SF}{SF}{spinless fermions}
\newacronym{fRG}{fRG}{functional renormalization group}
\newacronym{LE}{LE}{Luther\hyp Emery}
\newacronym{ASP}{ASP}{adiabatic state preparation}
\newacronym{VQE}{VQE}{variational quantum eigensolver}
\newacronym{METTS}{METTS}{minimally\hyp entangled typical thermal states}
\newacronym{SSH}{SSH}{Su\hyp Schrieffer\hyp Heeger}
\newacronym{GSE}{GSE}{Global Subspace Expansion}
\newacronym{LSE}{LSE}{Local Subspace Expansion}
\newacronym{LVO}{LVO}{LiV$_2$O$_4$}
\newacronym{QE}{QE}{\textsc{Quantum~ESPRESSO}}
\newacronym{MLWF}{MLWF}{maximally localized Wannier function}
\newacronym{W90}{W90}{\textsc{Wannier90}}
\newacronym{WF}{WF}{Wannier function}
\newacronym{BZ}{BZ}{Brillouin Zone}
\newacronym{DOS}{DOS}{density of states}
\newacronym{FL}{FL}{Fermi liquid}
\newcommand{\ascaddress}{Department of Physics and Arnold Sommerfeld Center for Theoretical Physics (ASC), Ludwig-Maximilians-Universit\"{a}t M\"{u}nchen, D-80333 Munich, Germany}
\newcommand{\mcqstaddress}{Munich Center for Quantum Science and Technology (MCQST), D-80799 M\"{u}nchen, Germany}

\newcommand{\CCQ}{Center for Computational Quantum Physics, Flatiron Institute, 162 5th Avenue, New York, NY 10010, USA}
\newcommand{\saclay}{Universit\'e Paris-Saclay, CNRS, CEA, Institut de Physique Th\'eorique, 91191, Gif-sur-Yvette, France}
\newcommand{\college}{Coll\`ege de France, 11 Place Marcelin Berthelot, 75005 Paris, France}
\newcommand{\cpht}{CPHT, CNRS, Ecole Polytechnique, IP Paris, F-91128 Palaiseau, France}
\newcommand{\unige}{
DQMP, University of Geneva, 24 quai Ernest-Ansermet, 1211 Geneva, Switzerland
}

\newcommand{\TFL}{$T_\mathrm{FL}\,$}

\newcommand{\Tonset}{$T_\mathrm{onset}\,$}

\newcommand{\JAF}[0]{J_{\mathrm{AF}}}

\newcommand{\aag}{\ensuremath{a_{1g}}}
\newcommand{\egpi}{\ensuremath{e_{g}^\pi}}

\definecolor{corn_flower}{HTML} {78A1E5}
\definecolor{lmugreen}{RGB}{0.0, 148, 64}
\definecolor{Gray}{gray}{0.9}
\newif\ifshowcomments\showcommentstrue
 
\definecolor{midnight_blue}{HTML} {002952}
\definecolor{hot_pink}{HTML} {D15D84}
\definecolor{corn_flower}{HTML} {78A1E5}
\definecolor{purple}{HTML} {4C3B55}
\definecolor{mauve}{HTML} {905171}
\definecolor{tiffany_blue}{HTML} {6AA4B0}
\definecolor{gunmetal_grey}{HTML} {4C5355}
\definecolor{honeysuckle}{HTML} {C44B4F}
\definecolor{blueish}{HTML} {226E9C}
\definecolor{bluegray}{HTML} {607D86}
\definecolor{cinnabar}{HTML} {E66354}
\definecolor{walnut}{HTML} {4D181C}
\definecolor{mahagony}{HTML} {4B1816}
\definecolor{aegean_blue}{HTML} {144058}
\definecolor{honey}{HTML} {E58D2E}
\definecolor{persimmon}{HTML} {DD671E}
\definecolor{mimosa}{HTML} {D7A449}
\definecolor{lilac}{HTML} {D5CAE4}

\definecolor{darkteal}{HTML} {3A6D80}
\definecolor{amber}{HTML} {F3CD53}
\definecolor{squash}{HTML} {D56729}
\definecolor{vermillion}{HTML} {9D402D}
\definecolor{cascades_green}{HTML} {355952}
\definecolor{baby_pink}{HTML} {FDC2E4}
\definecolor{wisteria}{HTML} {D3C5E5}
\definecolor{burgundy}{HTML} {800020}

\colorlet{colorUOpO}{midnight_blue}
\colorlet{colorUOpXXV}{persimmon}
\colorlet{colorUOpL}{burgundy}
\colorlet{colorUOpLXXV}{cascades_green}

\definecolor{orcidlogocol}{HTML}{A6CE39}

\begin{document}

\title{LiV$_2$O$_4$: Hund-Assisted Orbital-Selective Mottness}

\author{M.~Grundner\hspace{0.069cm}\orcidlink{0000-0002-7694-0053}}
\affiliation{\ascaddress}
\affiliation{\mcqstaddress}

\author{F.~B.~Kugler\hspace{0.069cm}\orcidlink{0000-0002-3108-6607}}
\affiliation{\CCQ}
\affiliation{Institute for Theoretical Physics, University of Cologne, 50937 Cologne, Germany}

\author{O.~Parcollet\hspace{0.069cm}\orcidlink{0000-0002-0389-2660}}
\affiliation{\CCQ}
\affiliation{\saclay}

\author{U.~Schollw\"ock\hspace{0.069cm}\orcidlink{0000-0002-2538-1802}}
\affiliation{\ascaddress}
\affiliation{\mcqstaddress}

\author{A.~Georges\hspace{0.069cm}\orcidlink{0000-0001-9479-9682}}
\affiliation{\college}
\affiliation{\CCQ}
\affiliation{\cpht}
\affiliation{\unige}

\author{A.~Hampel\hspace{0.069cm}\orcidlink{0000-0003-1041-8614}}
\email{mail@alexander-hampel.de}
\affiliation{\CCQ}

\date{\today}

\begin{abstract}

We show that the remarkably small Fermi-liquid coherence scale and large effective mass observed in LiV$_2$O$_4$ are due to the proximity of a Hund-assisted orbital-selective Mott phase (OSMP).
Our work is based on an ab initio dynamical mean-field approach, combining several quantum impurity solvers to capture the physics from high to very low temperature. 
We find that the Hund coupling plays a crucial role in rearranging 
the orbital populations and in generating the heavy mass and low coherence scale. 
The latter is found to be approximately 1--2 Kelvin, even though 
the most correlated orbital is found to be significantly doped ($\sim 10\%$) away from half-filling. 
A flat quasiparticle band appears near the Fermi level as a result of the strong 
electronic correlations. Finally, we discuss our results in comparison to experiments.

\end{abstract}

\maketitle

The `heavy fermion' (HF) phenomenon -- quasiparticles acquiring very large
effective masses -- is one of the most spectacular manifestations of strong
electronic correlations.  It is usually observed in materials having conduction
electrons hybridizing with very localized degrees of freedom, typically
$f$-orbitals in rare-earth or actinide compounds.  In this context, the
discovery of a remarkably large specific-heat enhancement in LiV$_2$O$_4$
(LVO), a transition-metal oxide involving no $f$-electrons, came as quite a
surprise~\cite{Kondo1997, Takagi1999,Urano2000}.  Indeed, only few other
transition-metal compounds displaying very large mass enhancements are known to
this day~\cite{Nidda2003,Crispino2023}.

\begin{figure}[t]
\centering
\includegraphics[width=\linewidth]{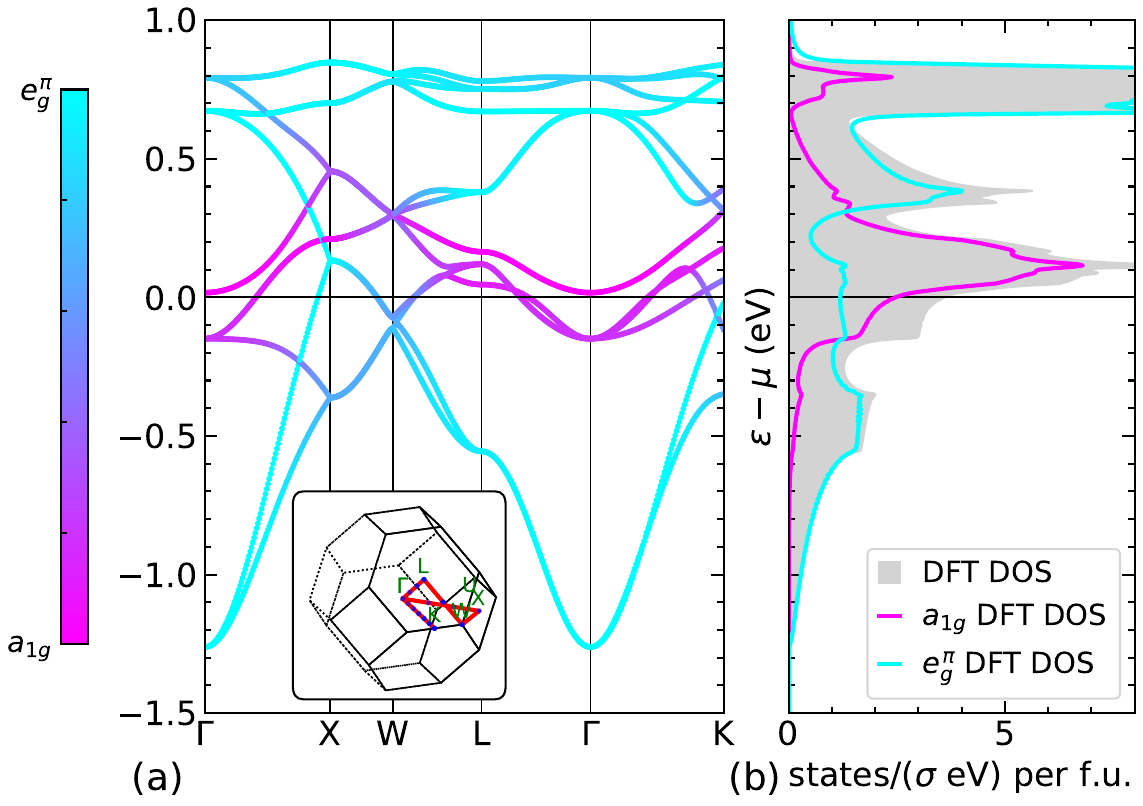}
\caption{(a) DFT band structure of \gls{LVO} for the four \aag{}- and eight \egpi{}-like bands around the Fermi level. 
The eigenvalues of the maximally localized Wannier functions (\glspl{MLWF}) perfectly match the DFT bands \cite{SM}.
Color: orbital character. 
Inset: Brillouin zone with high-symmetry points.
(b) DFT \gls{DOS} and projections onto the \glspl{MLWF}.} 
\label{fig:dft_results}
\end{figure}

The electronic structure of this material obtained with the density functional theory
(DFT) displays two sets of bands close to the Fermi level as shown in Fig.~\ref{fig:dft_results}~\cite{Matsuno1999,Singh1999}:
one set of \aag{} character, with a narrow bandwidth ($\sim\! 0.7$~eV), and one set of \egpi{} character with a broader bandwidth ($\sim\! 1.9$~eV).
A connection to HF materials was therefore suggested in early works~\cite{Anisimov1999,Singh1999,Varma_1999,Hopkinson2002}, 
with the \aag{} (resp.\ \egpi{}) band playing the role 
of localized states (resp.\ conduction electrons).
It was later realized, though, that this picture is not tenable. From a
theoretical viewpoint, the narrow band is too broad to play the role of
localized states, and, importantly, the coupling between the narrow and wide
band is dominantly an intra-atomic Hund coupling which favors spin alignment, not an antiferromagnetic Kondo
exchange~\cite{Arita2007,LeHur2007,Shimizu2012}. 
Furthermore, some experimental features are markedly different from conventional HFs. 
The onset of Kondo coherence in HFs is typically signaled by a coherence peak in the temperature ($T$) dependence of the resistivity, 
followed by a lower scale \TFL below which $\sim T^2$ Fermi-liquid (FL) behavior holds.  
While a small FL scale \TFL$\sim\! 2$~K is indeed observed in LVO, 
the resistivity, however, monotonously increases upon heating, eventually reaching a `bad metal' regime at high $T$~\cite{Takagi1999,Urano2000,Onoda2023}. 
From various experimental probes, the onset temperature of electronic coherence is estimated as
\Tonset$\sim$ 10--30~K~\cite{Takagi1999, Johnston1999,Urano2000}. 

Aware of the difficulties of the Kondo scenario, Arita et al.\  
proposed an alternative explanation~\cite{Arita2007}.  In their picture, all
the action takes place in the narrow \aag{} band, which is viewed as a doped
Mott insulator, while the broader \egpi{} bands are merely spectators.  In this description, a tiny hole doping ($\sim\! 2\%$) of the \aag{} band yields the observed large effective mass. 
In support of this picture, the authors of Ref.~\cite{Arita2007} performed
dynamical mean-field theory (DMFT)~\cite{Georges1996} calculations on a
simplified two-orbital model down to room temperature  
(as well as projective Monte Carlo calculations at $T\!=\! 0$). 
However, because of the very low energy scales involved, a full calculation of a
realistic model for LVO in the DFT+DMFT framework covering the whole $T$ range from above \Tonset to below \TFL has not been achieved yet~\cite{Nekrasov2003}.

In this Letter, we leverage on recent advances in 
DMFT impurity solvers to achieve this goal and elucidate the physics
responsible for the HF behaviour in LVO. 
To probe the full crossover from a higher-$T$ incoherent metal to the
low-$T$ heavy-mass FL and to access the low FL scale, 
we use a combination of three DMFT solvers:
Quantum Monte Carlo (QMC) for 10~K~$ \!<\! T \!<\! $~300~K, 
tensor networks (TN) for $T \!>\! $~1~K 
and the numerical renormalization group (NRG) for $T \!<\! $~1~K. 

We establish a connection to an orbital-selective Mott phase (OSMP)~\cite{Anisimov2002,Biermann_2005,deMedici_2009,Shimizu2012,Crispino2023,Samanta2024}, which is crucial in many systems, including iron-based superconductors~\cite{Herbrych2019,PhysRevLett.112.177001,PhysRevB.96.125110,PhysRevLett.132.136504,PhysRevLett.110.146402}, and show that LVO is close to a {\it Hund-assisted OSMP}. Here, the inter-orbital Hund coupling $J$ (and therefore the \egpi{} bands) plays a central role.
First, it induces a redistribution of the
(\aag{}, \egpi{}) orbital populations per vanadium due to interaction from
$(0.4, 2\times 0.55)$ in DFT to $(0.9, 2\times 0.3)$ in DMFT. Second, it is
responsible for the heavy mass and low \TFL{} via  the spin-blocking mechanism
characteristic of Hund systems in which electrons can only hop between atomic
configurations with maximum spin~\cite{Werner2008,Haule2009,Georges2013,GeorgesKotliar2024}. 
LVO is shown to be close to an OSMP in which the $a_{1g}$
orbital is prevented to fully localize at $T=0$ due to inter-orbital hopping
\cite{Kugler2022a}. Remarkably, we obtain a low \TFL{} already at $\sim\! 10\%$ doping of the \aag{} band, in
contrast to the $\sim\! 2\%$ doping of Ref.~\cite{Arita2007}.

{\it Electronic structure and effective model---}%
\gls{LVO} crystallizes in the fcc spinel structure (space group Fd3m).
Crystal structure data from neutron scattering are available down to 12~K~\cite{Kondo1997}. 
The primitive unit cell contains 14 atoms, with 4 V atoms, each embedded in an octahedral crystal field of the surrounding oxygen atoms. 
The octahedra themselves are corner-sharing and the local point group is trigonal $D_{3d}$, lowering the symmetry of the system and splitting the $t_{2g}$ states into a single \aag{} and a doubly degenerate \egpi{} orbital higher in energy.

We perform DFT calculations using the \textsc{Quantum~ESPRESSO}
software package fixing the structural parameters to the 12~K experimental values \cite{Kondo1997}.
Maximally localized Wannier functions (MLWFs) 
for the \aag{} and \egpi{} states are constructed with
\textsc{Wannier90}~\cite{Mostofi_et_al:2014},
accurately reproducing the DFT low-energy electronic structure. 
In Fig.~\ref{fig:dft_results}(a), the bandstructure of the Wannier Hamiltonian is shown together with its projection on the \aag{} and \egpi{} states. 
The projected density of states (DOS) in Fig.~\ref{fig:dft_results}(b) reveals that the \aag{} DOS is sharply peaked at 0.1~eV above the Fermi level, whereas the \egpi{} DOS is relatively flat. 

The constrained random phase approximation~\cite{Aryasetiawan2004} 
as implemented in the RESPACK code~\cite{respack2021} is used to determine the effective Coulomb interaction at low energy.
We focus on the static, $\omega \!=\! 0$ limit 
and fit the resulting four-index tensor to the symmetrized Kanamori form (including spin-flip and pair-hopping terms) with three independent parameters. The optimal fit yields $U \!=\! 3.94$~eV, $U' \!=\! 2.83$~eV, and $J \!=\! 0.56$~eV, which violates the full rotational invariance ($U'=U \!-\! 2J$) by only $\sim 0.01$~eV.

\begin{figure}[t]
\centering
\includegraphics[width=0.9\linewidth]{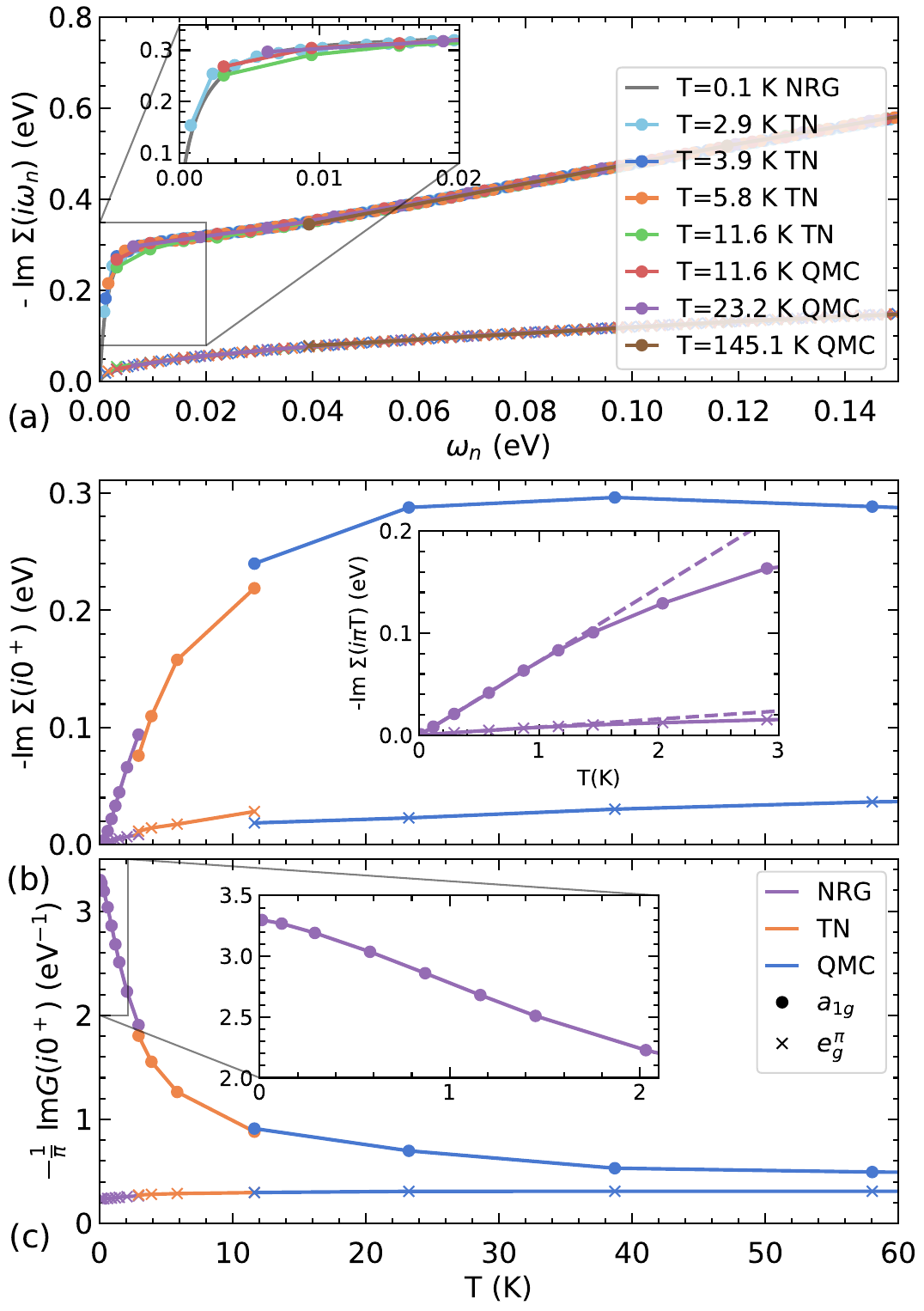}
\caption{(a) $\mathrm{Im}\Sigma(i\omega_n)$ for the \aag{} orbital (circles) and \egpi{} orbitals (crosses) at various $T$ from DFT+DMFT using QMC, TN, and NRG. 
Inset: zoom at low frequencies.
(b) Scattering rate and (c) spectral function at the Fermi level as a function of $T$.
Results at $i0^+$ in (b)--(c) for QMC and TN were obtained by fitting a fourth-order polynomial to the lowest eight Matsubara points.
The inset in (b) reveals the FL behavior of the self-energy, $-\mathrm{Im}\Sigma(i \pi T) \propto T$ \cite{Chubukov2012}, and the inset of panel (c) displays the saturation of the spectral function at low $T$.
}
\label{fig:dmft_sigma_Aw0}
\end{figure}

We use DMFT to solve the effective many-body model, with a combination of impurity solvers to cover all $T$ regimes (see \cite{SM} for details).
First, we use the continuous-time hybridization-expansion QMC solver 
implemented in the TRIQS software library~\cite{triqs,Seth2016} and its interface to electronic structure codes \textsc{TRIQS/DFTTools} and \textsc{TRIQS/solid$\_$dmft}~\cite{Aichhorn2016,Merkel2022,SM}.
QMC calculations were performed down to 11.6 K ($1/1000$ eV).
Second, the $T\!=\!0$ TN-based solver from \cite{Wolf2015,Linden2020,Karp2020} 
was extended to $T \!>\! 0$ using thermal-state purification \cite{Verstraete2004,Schollwoeck2011} and applied down to 2.9 K ($1/4000$ eV).
Third, to access the sub-$1\,\mathrm{K}$ regime, we use 
NRG, similarly as in Refs.~\cite{Kugler2020,Kugler2024}.
To make non-degenerate three-orbital NRG calculations tractable, we increase the local symmetry by neglecting 
the pair-hopping part of the interaction when using this solver~\cite{Kugler2020,SM}.

{\it Crossover to a heavy FL---}%
The emergence of the low-$T$ heavy FL is illustrated in 
Fig.~\ref{fig:dmft_sigma_Aw0}. Figure~\ref{fig:dmft_sigma_Aw0}(a) shows the imaginary part of the self-energy for each orbital as a function of Matsubara frequency for several $T$.
The \aag{} self-energy shows almost no $T$ dependence for $\omega_n > 0.01\,$eV. Only at very low $T$, a sudden drop
at low frequencies occurs (best visible in the inset) for calculations below the {\it onset temperature} \Tonset$\sim 12$~K.
This is a hallmark of strong correlations and 
corresponds to a sharp crossover from a high-$T$
incoherent regime with a large scattering rate to a low-$T$ coherent regime.
This is further supported by  
Fig.~\ref{fig:dmft_sigma_Aw0}(b), which shows the
extrapolated zero-frequency value of the \aag{} scattering rate
$-\mathrm{Im}\Sigma_{a_{1g}}(i0^+)$ as a function of $T$.
The inset of Fig.~\ref{fig:dmft_sigma_Aw0}(b) depicts $-\mathrm{Im}\Sigma_{a_{1g}}(i \pi T)$, from which we can infer that the FL is formed at 1--2~K when the plot becomes linear \cite{Chubukov2012}. Hence, we conclude that the $T$ range between 12~K, where the bending of $\mathrm{Im} \Sigma(i\omega_n)$ becomes visible, and 1~K can be identified as the FL crossover regime.
The local spectral function (DOS) at the Fermi level $A_{a_{1g}}(0)$ correspondingly increases, as shown in Fig.~\ref{fig:dmft_sigma_Aw0}(c). 
By contrast, the \egpi{} self-energy is smaller and depends more weakly on temperature and frequency.

The FL fully develops only below the {\it FL coherence temperature} \TFL$\sim$ 1--2~K, 
an order of magnitude smaller than \Tonset. 
The FL regime can also be characterized by $-\mathrm{Im}\Sigma \!\sim\! \omega^2 \!+\! \pi^2 T^2$ (see End Matter for $T\!=\! 0$) for the retarded self-energy and $\chi^{\prime\prime} \!\sim\! \omega$ for the imaginary part of the retarded local spin susceptibility (see End Matter).
The quasiparticle weights (from NRG at $T\!=\! 0$) are 
$Z_{\aag} \!\approx\! 0.003$ and  $Z_{\egpi} \!\approx\! 0.03$.

\begin{figure}[t]
\centering
\hspace{-0.5cm}\includegraphics[width=\linewidth]{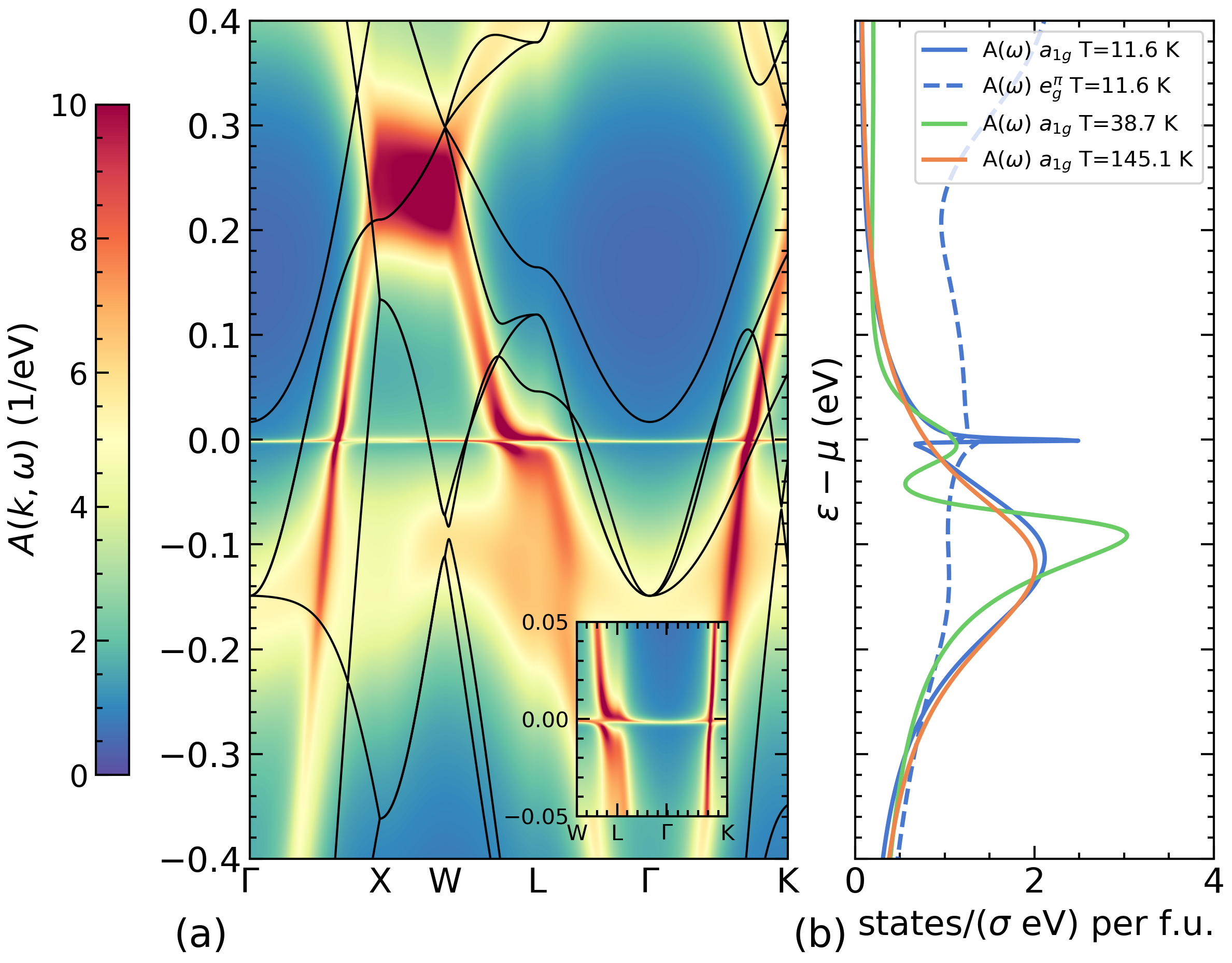}
\\
\includegraphics[width=\linewidth]{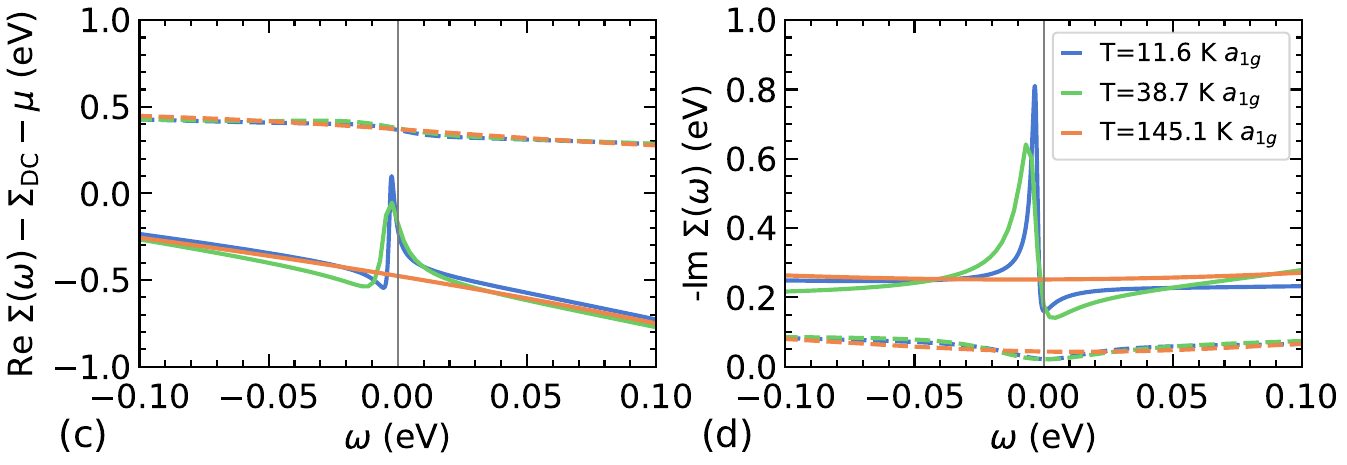}
\caption{(a) DFT+DMFT (QMC+Pad\'e) $k$-resolved spectral function at $T \!=\! 11.6$~K (black lines give the DFT bands).
Inset: zoom to the Fermi level, showing a flat quasiparticle band with suppressed spectral weight directly below $\omega\!=\!0$.
(b) DOS in the same energy range (see \cite{SM} for a wider range). The plain (dashed) blue line refers to the \aag{} (\egpi{}) orbital. The \aag{} result at larger $T$ is also shown.
(c) Real and (d) imaginary part of the QMC+Pad\'e real-frequency self-energies.}
\label{fig:akw_dos}
\end{figure}

{\it Quasiparticle bandstructure---}%
Figure~\ref{fig:akw_dos}(a) shows the momentum-resolved spectral function 
summed over orbitals at $T=11.6$~K.
The momentum-integrated 
spectral functions are displayed in panel (b) for several $T$. 
These results were obtained by 
analytic continuation of the QMC self-energies using Pad\'{e} extrapolation,
see Fig.~\ref{fig:akw_dos}(c)--(d),
and are consistent with NRG \cite{SM}. 

We see from Fig.~\ref{fig:akw_dos}(b) that the onset of coherence is associated with the growth of a remarkably narrow 
quasiparticle peak at the Fermi level, in line with the small $Z_{a_{1g}}$. 
This corresponds to the formation of a flat band near the Fermi level (see panel (a)),
together with the strong renormalization of the overall band structure
(compare the black lines from DFT).
The formation of the flat band can be understood from the fact that 
$\mathrm{Re}\Sigma_{a_{1g}}(\omega)$
(panel (c)) exhibits a very steep rise 
near $\omega \!=\! 0$. 
As a result, the quasiparticle equation (quoting for simplicity a one-band version) 
$\omega-\mathrm{Re}\Sigma(\omega)=\varepsilon_k-\mu$ has solutions $\omega_k$ which  
remain very close to zero energy for an extended range of momenta, as seen in panel (a) near the L-point. 

\begin{figure}[t]
\centering
\includegraphics[width=\linewidth]{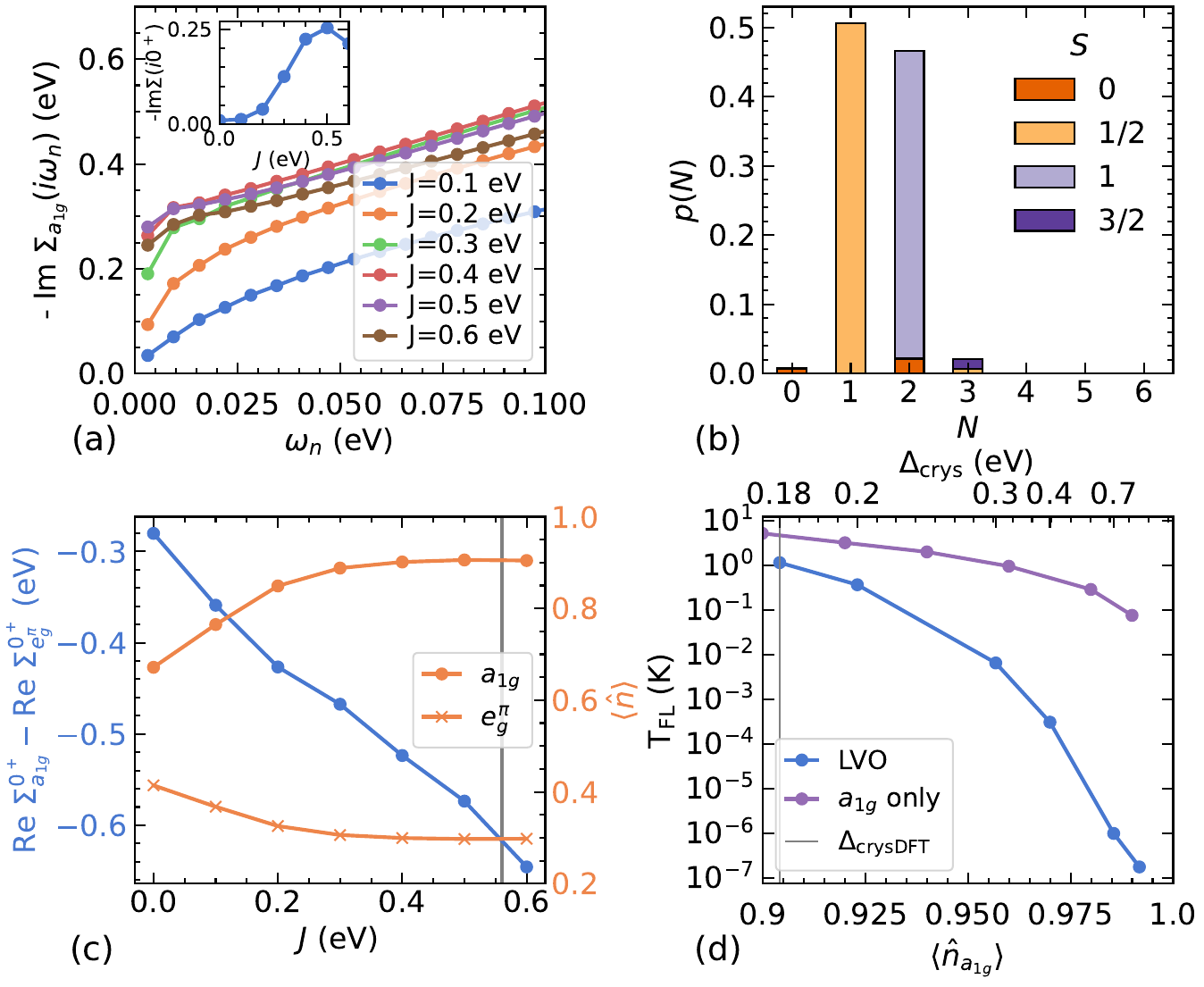}
\caption{(a) Evolution of $\Sigma_{a_{1g}}$ 
with increasing $J$ at $T \!=\! 11.6$~K.
Results are obtained with the TN solver at fixed $U \!=\! 3.94$~eV and $U' \!=\! U \!-\! 2J$. 
Inset: scattering rate obtained as in Fig.~\ref{fig:dmft_sigma_Aw0}(b).
(b) Histogram displaying the statistical weights of the relevant multiplets in each valence ($N$) and spin ($S$) sector, obtained from QMC at T=11.6~K.
(c) Evolution of the effective crystal field coming from $\mathrm{Re}\Sigma(0)$ and of the orbital populations with increasing $J$.
(d) \TFL for LVO as a function of the \aag{} occupancy when changing the crystal field (as indicated in the top horizontal scale, see text) and for an effective model involving only the \aag{} orbital, using the same intra-orbital $U$ value.
\TFL is estimated via the peak position of $\chi^{\prime\prime}(\omega)$ at $T \!=\! 0$ from NRG (see End Matter).
The vertical lines in panels (c) and (d) indicate the LVO values.}
\label{fig:hunds_coupling}
\end{figure}

{\it Discussion---}%
To elucidate the mechanism responsible for the strong correlations observed above, 
we first consider the effect of varying the Hund coupling $J$. 
Figure~\ref{fig:hunds_coupling}(a) shows the \aag{} self-energy on the Matsubara axis at $T \!=\! 11.6$~K for various $J$. 
At small $J$, the self-energies are small;
for $J \!\gtrsim\! 0.3$~eV, 
strong correlations develop. 
The self-energy becomes large and acquires the characteristic frequency dependence emphasized above 
(Fig.~\ref{fig:dmft_sigma_Aw0}(a)). 
This frequency dependence is reminiscent of Hund metals in the `spin freezing' 
regime~\cite{Werner2008,Haule2009,Kowalski2019} (see, e.g., Fig.~8 of Ref.~\cite{Kowalski2019}). 
Correspondingly, the extrapolated zero-frequency scattering rate at finite $T$ quickly increases as a function of $J$ 
(inset of panel Fig.~\ref{fig:hunds_coupling}(a)), indicating a suppressed coherence scale.
The Hund coupling is known to induce strong correlations through the `spin blocking' mechanism, which 
forces electrons to keep a high-spin configuration while hopping between different sites. 
The dominance of high-spin configurations is indeed supported by the valence- and spin-resolved 
histograms, obtained with QMC, see Fig.~\ref{fig:hunds_coupling}(b), 
or NRG \cite{SM}.  

In Fig.~\ref{fig:hunds_coupling}(c), we show the \aag{} and \egpi{} orbital occupancies as a function of $J$. 
The Hund coupling induces a significant redistribution of orbital populations~\cite{Nekrasov2003}, 
with the occupancy of the \aag{} orbital evolving from the DFT value $\sim\! 0.4$ to $\sim\! 0.9$ as $J$ is increased 
beyond $J \!\sim\! 0.3$~eV. 
Hence, the \aag{} orbital becomes $\sim\! 10\%$ hole-doped away from half-filling. This points at the relevance of 
Mott physics in this material, as proposed in Ref.~\cite{Arita2007}. 

To analyze the possible proximity of a Mott-insulating state, we perform a numerical experiment in which we add a crystal field to the Wannier Hamiltonian, while keeping the total $d$-electron count equal to 
1.5 by adjusting the chemical potential.
Figure~\ref{fig:hunds_coupling}(d) shows \TFL determined by NRG as a function of the \aag{} orbital occupancy (lower horizontal scale) or the applied crystal field (upper scale). 
We see that \TFL is driven to remarkably low values as half-filling of the \aag{} orbital is approached, with
values as low as $10^{-6}$~K at $1\%$ doping. 
This can be interpreted as the system reaching an OSMP~\cite{Anisimov2002,Biermann_2005,deMedici_2009,Shimizu2012,Crispino2023} in which the 
\egpi{} electrons remain itinerant while the \aag{} ones basically localize, full localization being prevented 
at $T \!=\! 0$ by inter-orbital hopping \cite{Kugler2022a}. 
The relevance of the OSM phenomenon and of Hund physics as a possible cause for 
$d$-electron HF behavior was recently emphasized by Crispino et al.~\cite{Crispino2023} and also previously mentioned qualitatively for LVO in the NMR study of Shimizu et al.~\cite{Shimizu2012}. 

While our results support the relevance of Mottness for LVO \cite{Arita2007}, 
they provide novel insights into the mechanism responsible for the HF behavior in LVO, 
namely the decisive role of Hund coupling and the wide range of permissible fillings.
In Fig.~\ref{fig:hunds_coupling}(d), we compare \TFL of LVO, as a function of orbital occupancy, 
to that of a single-band model with the same DOS as the one of the \aag{} orbital (also solved in DMFT).
We see that, in order to reach a value of \TFL in the range 1--2~K, as observed in experiments,  
a tiny doping of $\sim\! 1\%$ is required in the single-band picture. By contrast, the small \TFL is robustly generated from 
the realistic multi-orbital description, as a result of orbital and Hund physics, 
in a wide range of $a_{1g}$ doping. 
More broadly, our results emphasize that the multi-orbital character 
of the system is crucial, with inter-orbital correlations playing a key role, and hence that the 
\egpi{} bands are not merely spectators.

{\it Comparison to experiments---}%
Finally, we discuss how our results compare to experimental observations. 
The resistivity, Hall effect, specific heat and susceptibility measurements on single crystals~\cite{Urano2000}, as well as optical spectroscopy~\cite{Jonsson2007}, all point to electronic coherence setting in below \Tonset$\sim$ 10--30~K. Above \Tonset, LVO displays a large scattering rate and resistivity. 
This is consistent with our results in Fig.~\ref{fig:dmft_sigma_Aw0}. 
Furthermore, a $T^2$ dependence of the resistivity is only observed below \TFL $\!\sim\!$ 2~K, in excellent agreement with 
our estimate of \TFL $\!\sim\!$ 1--2~K. 
The gradual emergence below \Tonset of an extremely narrow quasiparticle peak reported in photoemission spectroscopy~\cite{Shimoyamada2006} also aligns with our Figs.~\ref{fig:dmft_sigma_Aw0}(c) and \ref{fig:akw_dos}(b). Angle-resolved photoemission experiments would be desirable to test the predictions of Fig.~\ref{fig:akw_dos}(a). 

The low-$T$ specific heat coefficient $C=\gamma T + \dots$ can be evaluated within DMFT 
from the zero-frequency spectral functions and quasiparticle weights of each orbital $m$ as $\gamma \!=\! \frac{2\pi^2}{3} k_B^2 \sum_{m} A_{m}(0)/Z_{m}$. 
The bare DFT value from our calculated DOS 
is $\gamma_\mathrm{DFT} \!=\! 17.1$~mJ/mol~K$^{-2}$, in good agreement with previous theoretical estimates~\cite{Nekrasov2003, Matsuno1999}.
Reported experimental values in the low-$T$ HF regime are in the range $\gamma/\gamma_\mathrm{DFT} \!\sim\! 25$~\cite{Kondo1997,Urano2000} 
to $\!\sim\! 30$~\cite{Kaps2001}. 
A comparable value is reached in our DMFT calculations at $T \!\sim\! 8$~K, but, at low $T$, we overestimate $\gamma/\gamma_\mathrm{DFT}$ 
by a large factor ($\sim\! 10$). 
We note, however, that no clear saturation of $C/T$ at low $T$ is seen experimentally, and that further increase of $\gamma$ at low $T$ 
was reported on some samples~\cite{Kaps2001,Okabe_2019}, so that the precise behavior of the 
specific heat of LVO at low $T$ is yet to be fully clarified. 

We conjecture that the overestimation of $\gamma$ at low $T$ comes from the intrinsic limitations 
of single-site DMFT. In this approach, $\gamma$ is inversely proportional to $Z$, which we find to be tiny for the \aag{} orbital. 
In a more accurate treatment beyond the single-site approximation, 
the inter-site magnetic exchange should intervene and reduce $C/T$ by 
reducing the entropy associated with fluctuating local moments above \Tonset. 
Indeed, slave-particle~\cite{kotliar_largeN_leshouches_1995,Florens_2004} or cluster-DMFT calculations~\cite{Parcollet_2004} 
yield, for a single band model, $\gamma/\gamma_\mathrm{DFT}\sim 1/(Z+\JAF/\epsilon_{\mathrm{F}})$ with 
$\JAF$ the antiferromagnetic super-exchange and $\epsilon_{\mathrm{F}}$ a typical (bare) electronic scale. 
In many oxides, the exchange term limits the value of $\gamma$ as compared to the DMFT $1/Z$ enhancement. 
Remarkably, this effect appears to be much less pronounced in LVO, allowing $\gamma$ to reach the 
large observed value. 
This is most likely due to frustration~\cite{Kaps2001,Burdin2002,Laad_2003,
Lee_neutrons_2001,Shimizu2012,Tomiyasu2014,Okabe_2019}
which results in a large amount of entropy stored at low $T$ and hence a 
reduced value of the exchange term $\JAF/\epsilon_\mathrm{F}$.  
NMR~\cite{Mahajan1998} and neutron scattering~\cite{Lee_neutrons_2001,Tomiyasu2014} experiments yield 
estimates of the effective exchange between V-atoms in the range $\sim$~20--30~K, comparable to \Tonset
and significantly smaller than typical values in unfrustrated oxides. 
A proper description of the spin correlations probed by these experiments 
likely requires a treatment beyond single-site DMFT \cite{SM}. 

{\it Outlook---}%
The physical picture emerging from our work is that LVO is close to a Hund-assisted 
OSMP. Recent experiments have indeed revealed that 
an insulating state can be induced by Li-intercalation~\cite{Yajima_2021} or, together with charge ordering, when 
subjecting this compound to uniaxial strain~\cite{Niemann2023}. 
While some qualitative aspects of this physical picture 
were anticipated before~\cite{Arita2007,Shimizu2012,Crispino2023},  
our work demonstrates how key progress on DMFT solvers now allows us to cover the full range of energy scales in 
this particularly challenging case and reliably establish the physical mechanism at hand. 
Our work also points at two necessary further developments: (i) including spatial fluctuations and computing 
spin and charge responses beyond single-site DMFT and (ii) exploring
instabilities toward charge and spin ordering 
dependent on external perturbations such as pressure, uniaxial strain, 
and chemical substitutions.

{\textit{Note added. }%
After completion of this paper, we became aware of the work subsequently posted on arXiv by \citeauthor{Backes:2025} which 
also addresses the physics of LiV$_2$O$_4$, with similar physical conclusions~\cite{Backes:2025}.

\acknowledgements
{\it Acknowledgements. }%
We acknowledge useful discussions and communications with Ryotaro Arita, Sophie Beck, Riccardo Comin, Andrea Damascelli, Luca de' Medici, Dongjin Oh, Dennis Huang, Xiangyu Luo, Giorgio Sangiovanni, Hidenori Takagi, Manish Verma, and Jan von Delft. 
M.G.\ and U.S.\ acknowledge support by the Deutsche Forschungsgemeinschaft (DFG, German Research Foundation) under Germany’s Excellence Strategy EXC-2111 No. 390814868. 
F.B.K.\ acknowledges funding from the Ministerium für Kultur
und Wissenschaft des Landes Nordrhein-Westfalen (NRW R\"uckkehrprogramm).
M.G.\ and U.S.\ are grateful to the Flatiron Institute and A.G.\ to the University of Munich for their hospitality. 
The Flatiron Institute is a division of the Simons Foundation.

{\it Data availability.} %
The data that support the findings of this 337
article are not publicly available. The data are available from 338
the authors upon reasonable request.

\bibliography{literature}

\clearpage

\onecolumngrid
\section{End Matter}
\twocolumngrid

{\it Appendix: Real-frequency dynamics---}%
We include, for completeness, several numerical results obtained directly in real frequencies from DFT+DMFT calculations using the NRG impurity solver.
Figure~\ref{efig:lin} shows the local spectral function and the real part of the self-energy of the $a_{1g}$ orbital on linear scales. 
The main panels give an overview at a large frequency range from $-4$~eV to $3$~eV.
The results from three choices of temperatures below $3$~K completely overlap.
Close to zero frequency, one notices a strikingly sharp low-energy feature.
The insets, magnifying this low-energy feature, reveal how the quasiparticle peak (top panel) builds up with decreasing temperature and the slope of $\mathrm{Re}\Sigma$ (bottom panel) increases, corresponding to a decreasing quasiparticle weight.

\vspace{0.5cm}

\begin{figure}[H]
\centering
\includegraphics[width=0.45\textwidth]{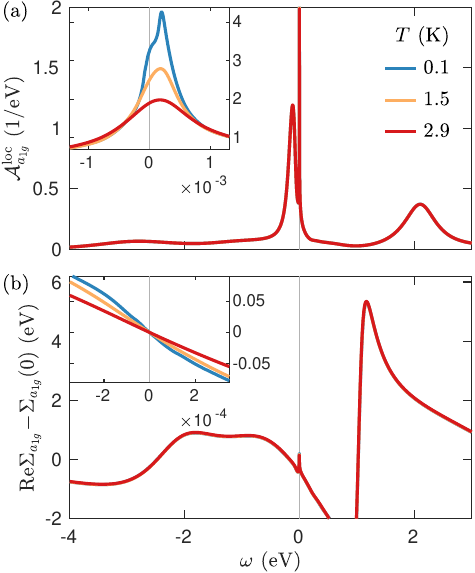}
\caption{(a) Local spectral function and (b) real part of the self-energy of the $a_{1g}$ orbital, for three choices of low temperature values, on linear scales.}
\label{efig:lin}
\end{figure}

\newpage 

Figure~\ref{efig:log} shows the imaginary parts of the $a_{1g}$ self-energy and of the local spin susceptibility on logarithmic scales, obtained at $T \!=\! 0$.
Colors indicate three choices of the crystal field;
$\Delta \!\approx\! 0.18$ is the appropriate value for LVO,
and larger $\Delta$ is used to approach an OSM state.
The top panel demonstrates the large scattering rate in the incoherent regime at finite energies
and the crossover to FL behavior $-\mathrm{Im}\Sigma \!\sim\! \omega^2$ at extremely low frequencies.
The bottom panel supports this picture:
The incoherent regime is characterized by an extremely large spin susceptibility, while,
in the FL regime, $\chi^{\prime\prime} \!\sim\! \omega$ decreases as energy is decreased.
The peak position of $\chi^{\prime\prime}$, $\omega_{\mathrm{max}}$, can be used to estimate $T_\mathrm{FL}$ (e.g., $T_{\mathrm{FL}} \!=\! \omega_{\mathrm{max}}/2$).
The result for LVO ($\Delta \!\approx\! 0.18$) is roughly $0.1$~meV or $1$~K.

\vspace{0.6cm}

\begin{figure}[H]
\centering
\includegraphics[width=0.45\textwidth]{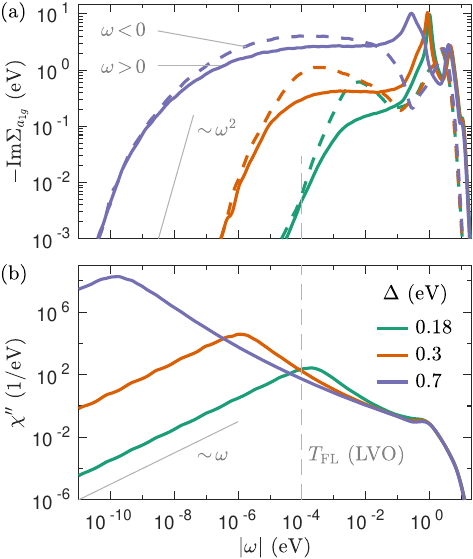}
\caption{Imaginary parts of (a) the self-energy of the $a_{1g}$ orbital and (b) the local spin susceptibility, for three choices of crystal-field splitting, on logarithmic scales.}
\label{efig:log}
\end{figure}
\end{document}


\title{Supplemental Material \\ LiV$_2$O$_4$: Hund-Assisted Orbital-Selective Mottness}

\author{M.~Grundner\hspace{0.069cm}\orcidlink{0000-0002-7694-0053}}
\affiliation{\ascaddress}
\affiliation{\mcqstaddress}

\author{F.~B.~Kugler\hspace{0.069cm}\orcidlink{0000-0002-3108-6607}}
\affiliation{\CCQ}
\affiliation{Institute for Theoretical Physics, University of Cologne, 50937 Cologne, Germany}

\author{O.~Parcollet\hspace{0.069cm}\orcidlink{0000-0002-0389-2660}}
\affiliation{\CCQ}
\affiliation{\saclay}

\author{U.~Schollw\"ock\hspace{0.069cm}\orcidlink{0000-0002-2538-1802}}
\affiliation{\ascaddress}
\affiliation{\mcqstaddress}

\author{A.~Georges\hspace{0.069cm}\orcidlink{0000-0001-9479-9682}}
\affiliation{\college}
\affiliation{\CCQ}
\affiliation{\cpht}
\affiliation{\unige}

\author{A.~Hampel\hspace{0.069cm}\orcidlink{0000-0003-1041-8614}}
\email{ahampel@flatironinstitute.org}
\affiliation{\CCQ}

\date{\today}

\maketitle

In this Suppplemental Material, we give computational details of all employed methods,
DFT and cRPA, QMC, TN, and NRG,
and finally discuss magnetic order in LVO.

\section{DFT and low-energy model} \label{supp:dftlowenergy}

\begin{figure*}
\centering
a)\includegraphics[width=0.25\textwidth]{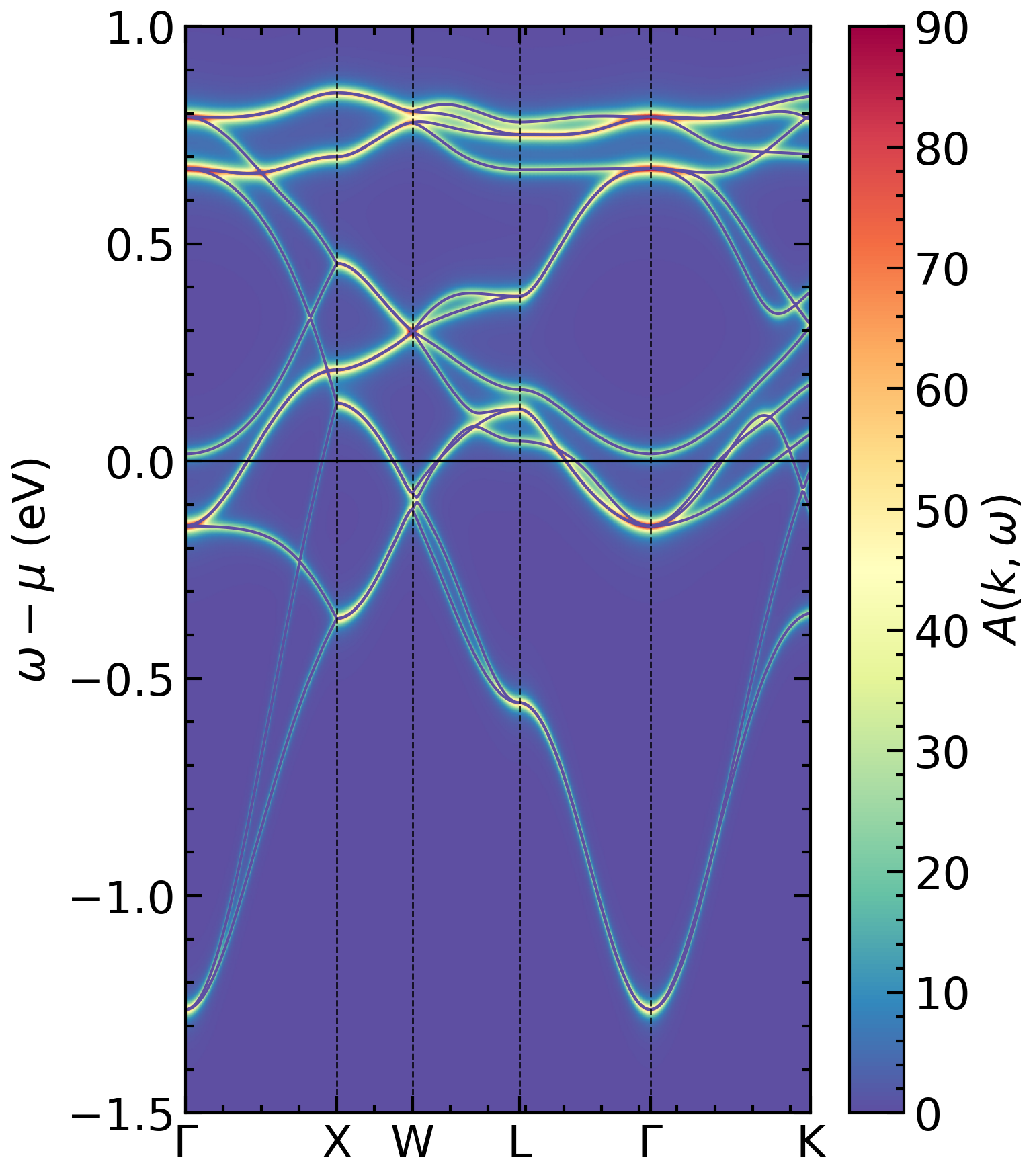}
b)\includegraphics[width=0.25\textwidth]{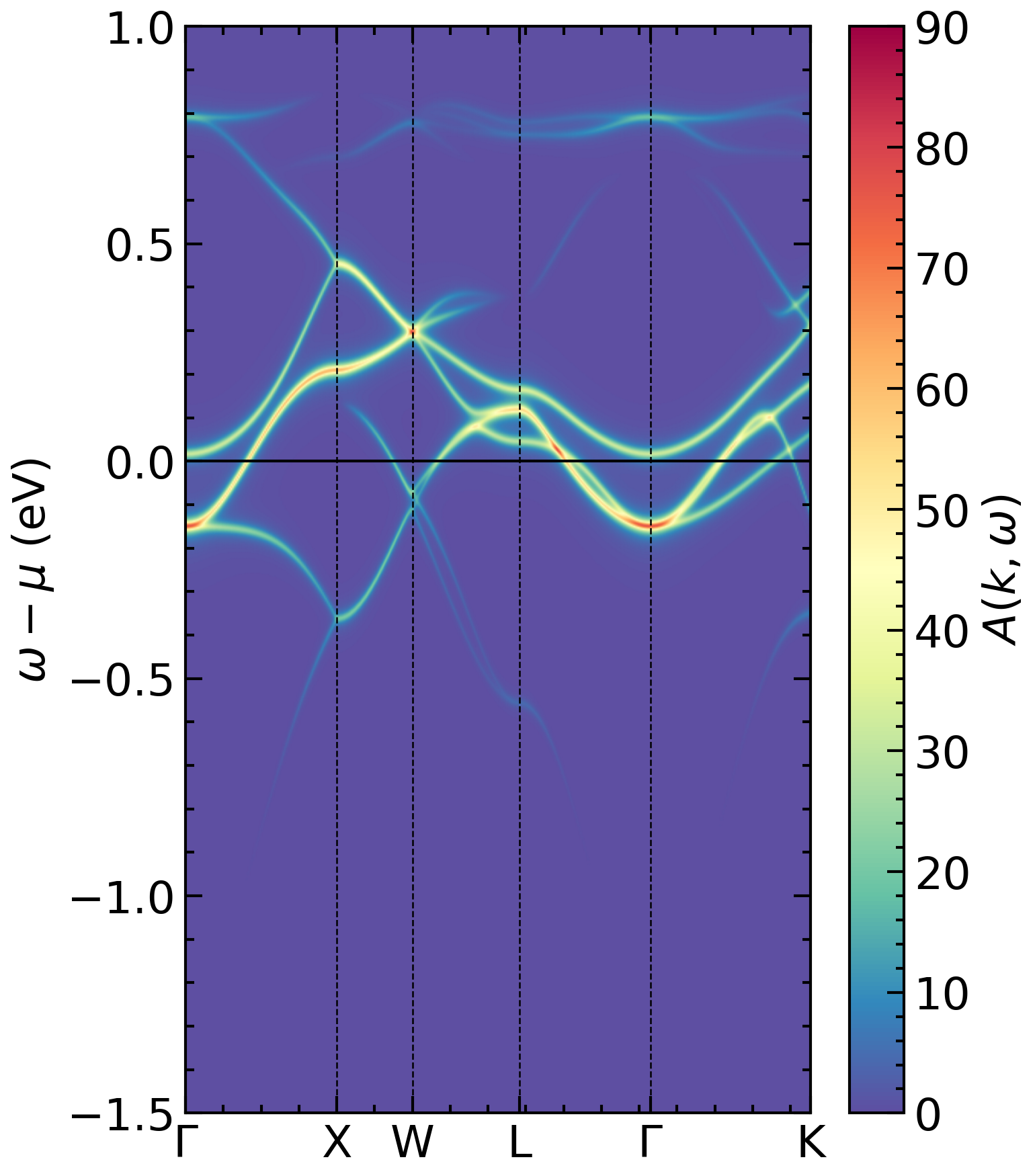}
c)\includegraphics[width=0.25\textwidth]{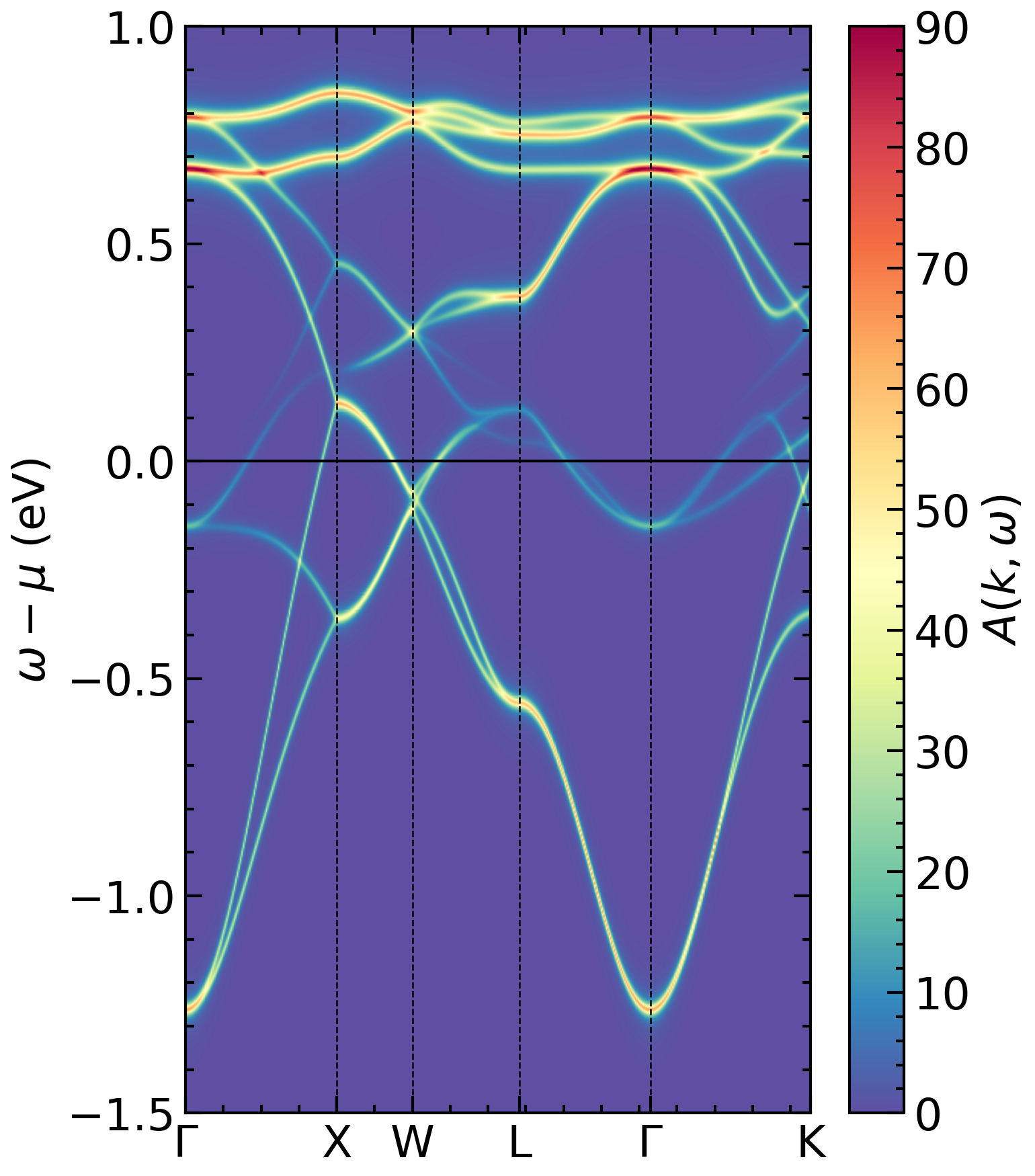}
\caption{DFT band structure and DOS calculated by wannier90 for the 12 orbital low-energy model for the \aag and \egpi bands around the Fermi level. (a) band structure calculated from the wannier90 model Hamiltonian (gray lines), and the corresponding non interacting spectral function with small broadening applied. (b) Spectral function projection on the \aag{} orbital on each of the four V-sites. (c) Projection on the two \egpi{} orbitals on each V-site.}
\label{sfig:dft_bands_w90}
\end{figure*}

\Gls{DFT} calculations are performed using the \gls{QE} software package fixing the structure parameters to the experimental values reported for 12~K $a_0$=\SI{8.22694(3)}{\angstrom} and $x_\text{O}=0.26109(2)$~\cite{Kondo1997}. We performed calculations using the \textsc{ONCVPSP} norm-conserving pseudo-potentials~\cite{Hamann2013,vanSetten2018} in conjunction with the Perdew--Burke--Ernzerhof (PBE) exchange-correlation functional. We use a wavefunction cut-off of 90~Ry, and a density cut-off of 360~Ry, resulting in an energy error of \SI{<1}{\milli \electronvolt} per formula unit. A mesh of $11 \times 11 \times 11$ $k$-points has been chosen, matching this total energy error requirement. 

Further, we construct \glspl{MLWF} using \gls{W90}~\cite{Mostofi_et_al:2014} for the low-energy states around the Fermi level with dominantly V $d$-character.
This allows us to define a low-energy Hamiltonian $H^\text{W90}(\mathcal{R})$ in real space, capturing the essential physics of the system realistically. Furthermore, we rotate the Hamiltonian on each V-site into the crystal-field basis to diagonalize both the local non-interacting Hamiltonian and the hybridization function. 
For high accuracy low-temperature calculations, we leverage Wannier interpolation, Fourier transforming $H^\text{W90}(\mathcal{R})$ to a dense $41 \times 41 \times 41$ $k$-point mesh to avoid any $k$-discretization error. We followed this procedure for both an effective 12 band model (\aag{} plus \egpi{} orbitals) and an effective 20 band model (all five $d$-orbitals per V site).

We used the constrained random phase approximation (cRPA)~\cite{Aryasetiawan2004} as implemented in the RESPACK code~\cite{respack2021} to determine the effective Coulomb interaction for our low-energy model. By separating the electronic structure into a subspace near the Fermi level and the rest of the system we calculate the effective partially screened Coulomb interaction that can be identified as the effective interaction in our low-energy model. Formally, this means the separation of the total electronic polarizability $P = P_{\text{sub}} + P_{\text{rest}}$ where $P_{\text{sub}}$ is the polarizability for the correlated subspace and $P_{\text{rest}}$ is for the rest of the system. Then, the screened Coulomb tensor can be calculated in a local basis from the bare Coulomb interaction tensor $\bm{V}$, as $\bm{U}(\omega) = \bm{V}/[1 - \bm{V}P_{\text{rest}}(\omega)]$. We used the same well-localized correlated subspace basis for cRPA as used within DMFT, i.e., contributions to the polarizability for the target space are removed via their overlap with the Kohn--Sham states~\cite{Friedrich2011}, and the screened and bare Coulomb integrals are then evaluated in the maximally localized Wannier orbitals basis set of the correlated subspace treated in DMFT. Here, we limit ourselves to the static limit $\bm{U}(\omega = 0)$ of the screened interaction. We further post-process $\bm{U}(\omega = 0)$ to obtain a symmetrized Kanamori form of the interaction, which is easier to treat by the impurity solvers.

The Kanamori interaction Hamiltonian has the form
\begin{align}
\begin{split}\label{seqn:hubbard_kanamori}
  \hat{H}_U &= \frac{1}{2} \sum_{\sigma} \sum_{i} U \ \hat{n}_{i \sigma} \hat{n}_{i \bar{\sigma}} \\
  &+ \frac{1}{2} \sum_{\sigma} \sum_{i \neq j} \left[ U' \ \hat{n}_{i \sigma} \hat{n}_{j \bar{\sigma}} + ( U' - J) \hat{n}_{i \sigma} \hat{n}_{j \sigma} \right] \\
  &+ \frac{1}{2} \sum_{\sigma} \sum_{i \neq j} \left[ J \ c_{i \sigma}^\dagger c_{j \bar{\sigma}}^\dagger c_{i \bar{\sigma}} c_{j \sigma}  + J \ c_{i \sigma}^\dagger c_{i \bar{\sigma}}^\dagger c_{j \bar{\sigma}} c_{j \sigma} \right] \ .
\end{split}
\end{align}
Here, $i,j$ mark different orbitals, but the same side, $\sigma$ marks spin, and $\hat{n}=c^\dagger c$ is the density operator.
We solve the interacting many-body model in a one-shot fasion, i.e., without charge density updates to the DFT code.
DMFT calculations (except those with NRG as impurity solver) are performed utilizing the \textsc{TRIQS} software library~\cite{triqs}, and its interface to electronic structure codes \textsc{TRIQS/DFTTools}~\cite{Aichhorn2016}.

\section{Quantum Monte Carlo}
\label{sec:suppmatqmc}

\begin{figure}
    \centering
    \includegraphics[width=0.45\textwidth]{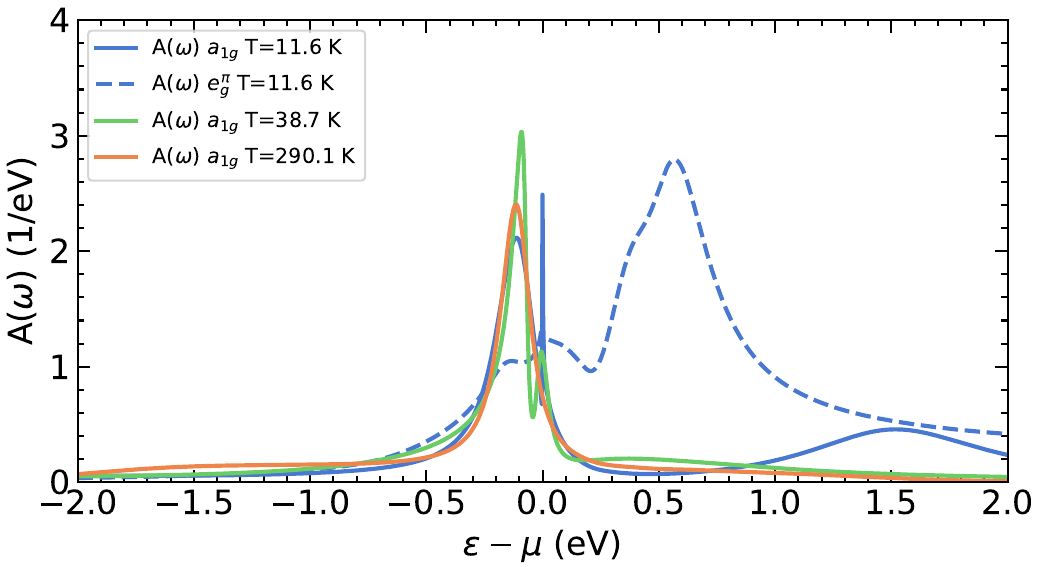}
    \caption{DOS from DMFT (QMC) by analytically continuing the Matsubara self-energy and integrating the real-frequency lattice Green's function. The solid (dashed) blue line refers to the \aag{} (\egpi{}) orbital at $T \!=\! 11.6$~K. 
    The \aag{} result at larger $T$ is also shown. Results are identical to Fig.~3(b) of the main text, but shown at larger energy range.
    }
    \label{sfig:akw_large_window}
\end{figure}

We employ \gls{QMC}-based solvers in the hybridization-expansion formulation using \textsc{TRIQS/cthyb}~\cite{Seth2016} and  
\textsc{triqs/solid\_dmft}~\cite{Merkel2022}. The solvers take into account all off-diagonal and also non density-density terms. Typically, the DMFT cycle is converged so that $|G_\text{imp} - G_\text{loc}|<10^{-4}$, but we additionally monitored explicitly the convergence of the impurity self-energy at the first few Matsubara frequencies to reliable investigate the low-energy physics. At low energies we required that the change in self-energy values between DMFT iterations is $\sim 1$~meV. \gls{QMC} calculations were performed from $290$~K down to $11.6$~K. To achieve the aforementioned accuracy, $5 \times 10^7$ independent measurements of $G_\text{imp}$ are performed in the final \gls{DMFT} iteration, the number of Matsubara frequencies is increased to up to $4 \times 10^4$ during the DMFT step at low temperatures, and the number of $\tau$ points is increased to $4 \times 10^5$ when measuring the fast decaying impurity Green function in \textsc{TRIQS/cthyb}. At very low temperatures, the warm-up time to thermalize \gls{QMC} is substantial in this system, with the chosen interaction parameters, and up-to $10^5$ warm-up cycles on each MC walker are necessary to thermalize and remove any spurious behavior in $\Sigma_\text{imp}$, which takes around 24 hours per impurity solver call.
High-frequency noise in the impurity self-energy is suppressed by fitting the tail of $\Sigma_\text{imp}$ up to fourth order in a window with moderate noise. The first two moments of the tail expansion, i.e., the Hartree shift Re$\Sigma_\text{imp}(i\omega_n \to \infty)$ and the leading order decay of Im $\Sigma_\text{imp}(i\omega_n)$, are measured to high precision directly in \textsc{TRIQS/cthyb}~\cite{labollita2023stabilizing} and fixed in the fitting, stabilizing the DMFT convergence.

In Fig.~\ref{sfig:akw_large_window}, we show the local spectral function (DOS) obtained from QMC in a larger energy window range compared to Fig.~3(b) in the main text. An upper Hubbard band of the \aag{} orbital can be observed at around 1.5~eV.

\section{Tensor Networks}
\label{supp:details_tns}
We use a tensor network solver for the DMFT impurity problem operating on the Matsubara axis~\cite{Wolf2015,Karp2020}. Green's functions are obtained via time evolution in imaginary time and subsequent Fourier transformation~\cite{Wolf2015}.
Calculations are performed using matrix product states (MPS) as implemented in the \textsc{SyTen} tensor network library~\cite{Syten}. 

The impurity interaction terms are given by the Hubbard--Kanamori Hamiltonian as defined in equation~(\ref{seqn:hubbard_kanamori}) with interaction parameters as determined above. Onsite energies and impurity\hyp bath hoppings are obtained from fitting the hybridization function \cite{Caffarel1994,Wolf2015,Linden2020}. Calculations were performed using matrix product states (MPS) with the three impurity orbitals $I_1$, $I_2$, and $I_3$ grouped together with their associated bath orbitals $B_1$, $B_2$, and $B_3$ kept in vicinity. Results in the main text are shown for $L_b=10$ bath sites per impurity orbital. We exploited all 5 symmetries of the Hamiltonian, namely particle number conservation ($U(1)$), spin degeneracy ($SU(2)$), and band parity conservation ($Z_2$) for each band.

\begin{figure}[t]
\centering
\includegraphics[width=0.45\textwidth]{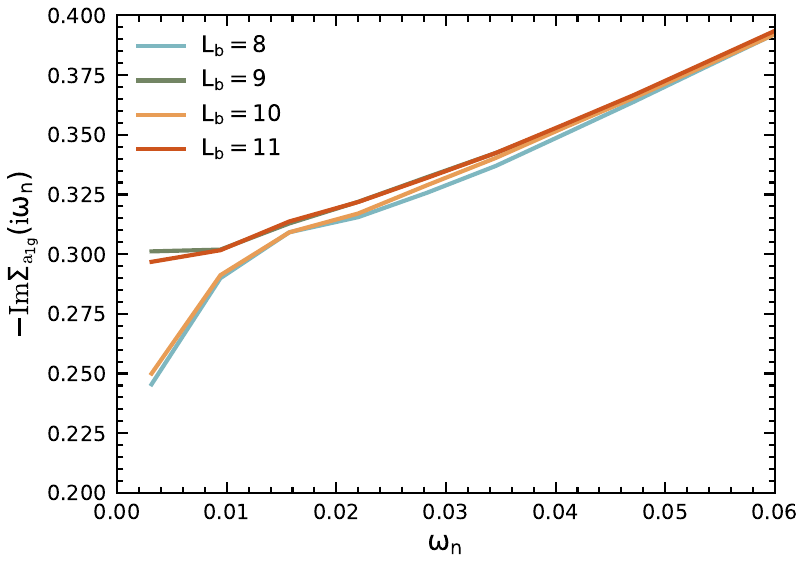}
\caption{Odd\hyp even effect in bath size of TN\hyp based simulations at $\beta=1000$. The even bath sizes give the physical result.}
\label{sfig:chi_T_dependency}
\end{figure}

We extended our zero\hyp temperature impurity solver~\cite{Wolf2015,Linden2020,Karp2020} using the purification technique~\cite{Verstraete2004,Schollwoeck2011}. Calculations were performed in a temperature range from a few hundred K down to 2.9 K ($\beta=4000$~eV$^{-1}$). In these calculations, the physical thermal state $\rho_P(\beta)$ is represented by tracing over the auxiliary degrees of freedom of a pure state $|\psi_\beta\rangle$ living in a Hilbert space of physical ($P$) and auxiliary ($A$) degrees of freedom
\begin{align}
    \rho_{P} (\beta) &= \Tr_{A}\ket{\psi_\beta}\bra{\psi_\beta} 
\end{align}
where we get $\ket{\psi_\beta}$ by an imaginary time-evolution
\begin{equation}
    \ket{\psi_\beta}=(\textrm{e}^{-\beta\hat{H}_P/2} \otimes \hat{I}_A) \ket{\psi_0}. 
\end{equation}
Here, $\hat{H}_P$ is the actual physical Hamiltonian, and $\hat{I}_A$ the identity acting on the auxiliary degrees of freedom. The requirement on $\ket{\psi_0}$ is that $\Tr_A \ket{\psi_0}\bra{\psi_0}=\hat{I}_P$, the (unnormalized) infinite-temperature thermal state.
All thermal expectation values can then be expressed as pure state expectation values $\bra{\psi_\beta} \cdots \ket{\psi_\beta}/Z_\beta$  with $Z_\beta= \langle \psi_\beta | \psi_\beta\rangle$. Here, we choose as $\ket{\psi_0}$ the product state of maximally entangled states on pairs of physical and ancilla orbitals given by 
\begin{equation}
|\psi_0^i\rangle = \frac{1}{2} (|0,\uparrow\downarrow\rangle + |\uparrow,\downarrow\rangle + |\downarrow,\uparrow\rangle + |\uparrow\downarrow,0\rangle),
\end{equation}
expressed in SU(2) invariant form. Due to the high accuracy required in the imaginary time evolution,
we used a combination of several time evolution techniques, while keeping a maximum of 1536 $SU(2)$ states (corresponding to about 5000 $U(1)$ states) and using a truncated weight of $w_t=10^{-11}$ throughout. We used the global Krylov method \cite{Paeckel2019} up to imaginary time $\tau=0.5$~eV$^{-1}$, the global subspace expansion method (GSE) \cite{Yang_2020prb} up to $\tau=5$~eV$^{-1}$ followed by two\hyp site TDVP \cite{Haegeman2011,Haegeman2016} up to $\tau=20$~eV$^{-1}$ and the local subspace expansion method (LSE) \cite{Yang_2020prb, Grundner2023} up to $\tau=\beta/2$. We found it helpful to use exponentially increasing time steps $\delta \tau_n=\min(0.1\times1.01^{n},0.5)$~eV$^{-1}$ during thermal state preparation to limit the impact of truncation and projection errors.

During the imaginary time evolution of our excited states for the calculation of the impurity Green's functions we again used a truncated weight of $w_t=10^{-11}$ and allowed for a maximum of 1536 $SU(2)$ states. Here, we used GSE up to $\tau=2$~eV$^{-1}$ followed by two\hyp site TDVP \cite{Haegeman2011,Haegeman2016} up to $\tau=25$~eV$^{-1}$ and LSE for the remainder of the time evolution up to $\tau=\beta/2$. Using either the GSE method or the global Krylov method for at least a single time step in the beginning is essential to ensure correct results, due to the inability of two\hyp site TDVP to produce the necessary basis states immediately after a local excitation~\cite{Yang_2020prb,Paeckel2019,Bertrand2021}.

As an example of the resulting accuracy, we consider the occupancy of the degenerate bands at the very low temperature $\beta=4000$~eV$^{-1}$, which is given by $0.29872814$ and $0.29875236$ respectively, with a deviation of $2.4 \times 10^{-5}$. To improve the numerical accuracy of the calculation of self-energies we used an improved estimator presented in \cite{Bulla1998}.

\begin{figure}[t]
\centering
\includegraphics[width=0.45\textwidth]{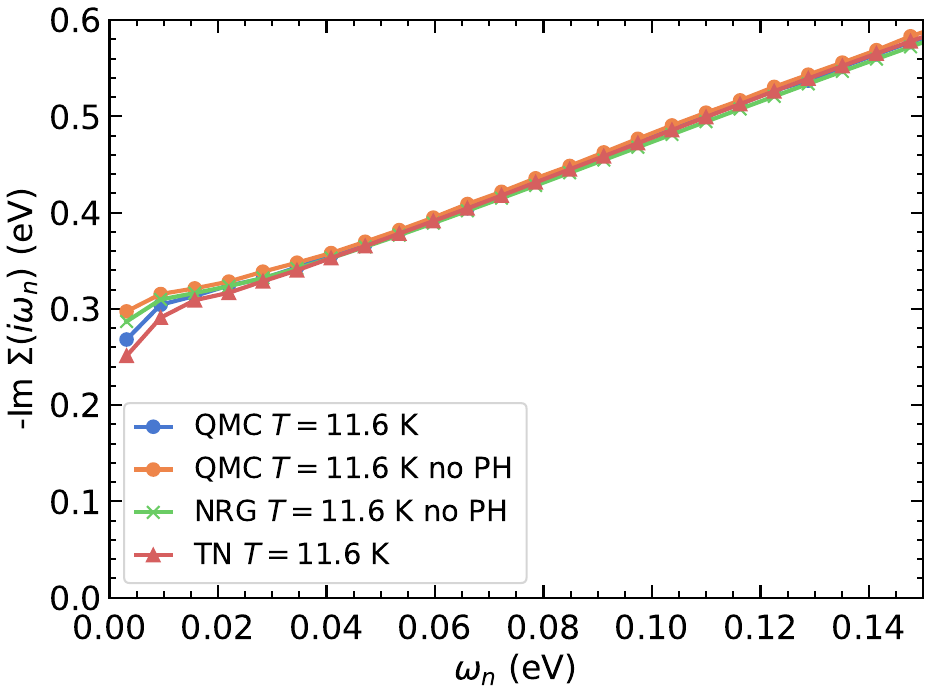}
\caption{Effect of pair hopping on the Matsubara self-energy.}
\label{sfig:pairhop}
\end{figure}

At the lowest temperatures, the TN calculations produce a low\hyp energy spectrum which depends on the parity of the bath sizes (and the resulting particle numbers) beyond expectable finite-size effects. This odd-even effect in particle numbers originates from the $a_{1g}$ orbital where there is one bath site with an on-site energy very close to zero (at $\beta=1000$, $-0.0035$~eV for $L_b=9$ and $-0.00087$~eV for $L_b=10$) which has an occupation by one particle more or less. At the same time, the differences in the chemical potentials between $L_b=9$ and $L_b=10$ are about $0.0039$~eV. The differences in the hybridization functions $\Delta_{a_{1g}}$ for $L_b=9$ and $L_b=10$ are notable only for the first two Matsubara frequencies for this $\beta$ and of order $\mathcal{O}(10^{-3})$ and $\mathcal{O}(10^{-4})$ respectively. The crux is that these respective energy scales to be resolved are at the limit of the currently possible. 

The physical choice that the correct results are obtained for the {\em even} bath sizes (with total spin $S=0$ in $T=0$ calculations which we also carried out) rests theoretically on the handshake with the QMC results at $\beta=1000$ and with NRG results at $\beta=4000$. Additionally, it is supported by the experimental knowledge that \gls{LVO} is a Fermi liquid at very low $T$, indicating that total spin $S=0$ prevails at very low $T$.

\section{Numerical Renormalization Group}
\label{supp:nrg}

We use the full density-matrix NRG \cite{Peters2006,Weichselbaum2007} in an implementation \cite{Lee2021} based on the QSpace tensor library \cite{Weichselbaum2012a,*Weichselbaum2012b,*Weichselbaum2020}. 
We employ $z$-averaging \cite{Zitko2009}, adaptive broadening \cite{Lee2016,Lee2017}, a symmetric improved self-energy estimator \cite{Kugler2022} and interleave the Wilson chains \cite{Mitchell2014,Stadler2016}. 
The number of $z$ shifts is $n_z = 3$;
the NRG discretization parameter is $\Lambda = 6$;
we interleave the three Wilson chains (one chain per orbital)
and keep up to $N_{\mathrm{keep}}=4 \times 10^5$ SU(2) multiplets during the iterative diagonalization.

\begin{figure}[t]
\centering
\includegraphics[width=0.45\textwidth]{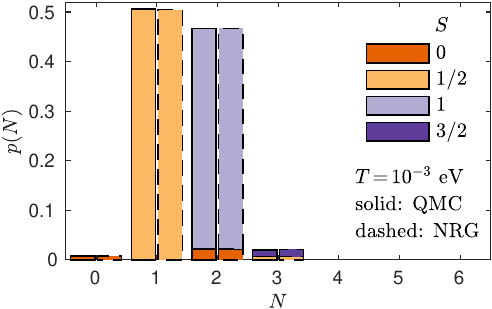}
\caption{Valence histogram obtained from QMC and NRG.}
\label{sfig:hist}
\end{figure}

Similarly as in Refs.~\cite{Kugler2020,Kugler2024}, we neglect the pair-hopping part of the Kanamori Hamiltonian, i.e., the very last term in Eq.~\eqref{seqn:hubbard_kanamori}.
Thereby, the charge per orbital is conserved, and the local symmetry is increased from $U(1) \otimes SU(2) \otimes (Z_2)^3$
to $SU(2) \otimes (U(1))^3$.
This simplification makes the NRG calculations considerably more efficient 
and was found to have only mild impact on the low-energy physics in Ref.~\cite{Kugler2020}.
In Fig.~\ref{sfig:pairhop}, we compare results with (QMC and TN) and without (QMC and NRG) pair hopping in the present context and again find that the effect is only mild.
Furthermore, Fig.~\ref{sfig:hist} shows the valence histogram,
i.e., the probability of different impurity configurations.
The results between QMC and NRG at $T \!=\! 1/1000$~eV agree perfectly.
NRG results at lower $T$ differ from those at $T \!=\! 1/1000$~eV only marginally.

\section{Magnetic correlations and ordering in DMFT}
\label{supp:magnetic order}

\begin{figure}[H]
\centering
\includegraphics[width=0.45\textwidth]{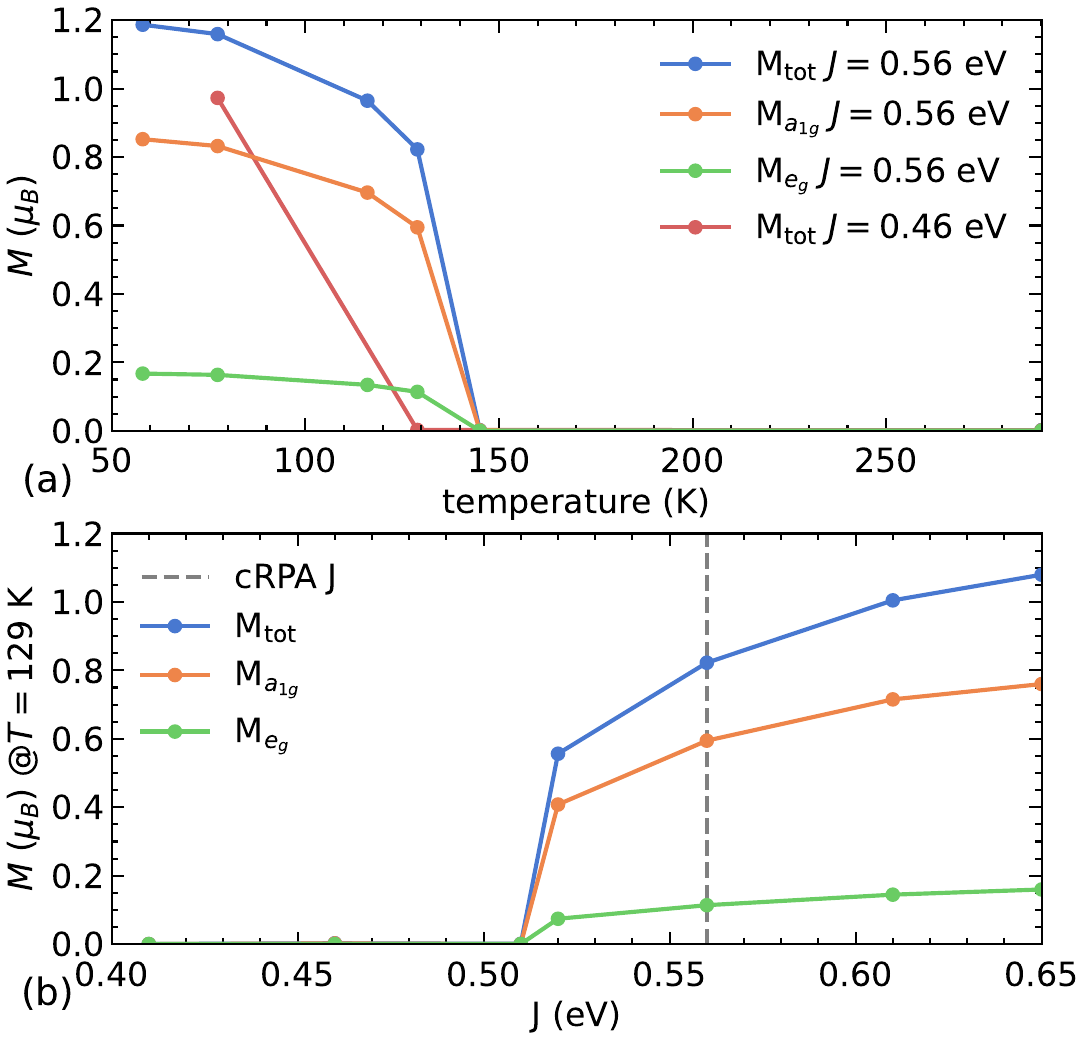}
\caption{(a) Orbital-resolved magnetic moments from DMFT (calculated via QMC) at fixed $U \!=\! 3.94$~eV and $U' \!=\! 2.83$~eV. For $J \!=\! 0.56$~eV, the system orders FM in single-site DMFT with a strong local moment in the \aag{} orbitals. 
(b) Magnetic moment calculated via QMC at $T \!=\! 129$~K and varying $J$.}
\label{sfig:magnetism}
\end{figure}

\begin{figure}[H]
\centering
\includegraphics[width=0.45\textwidth]{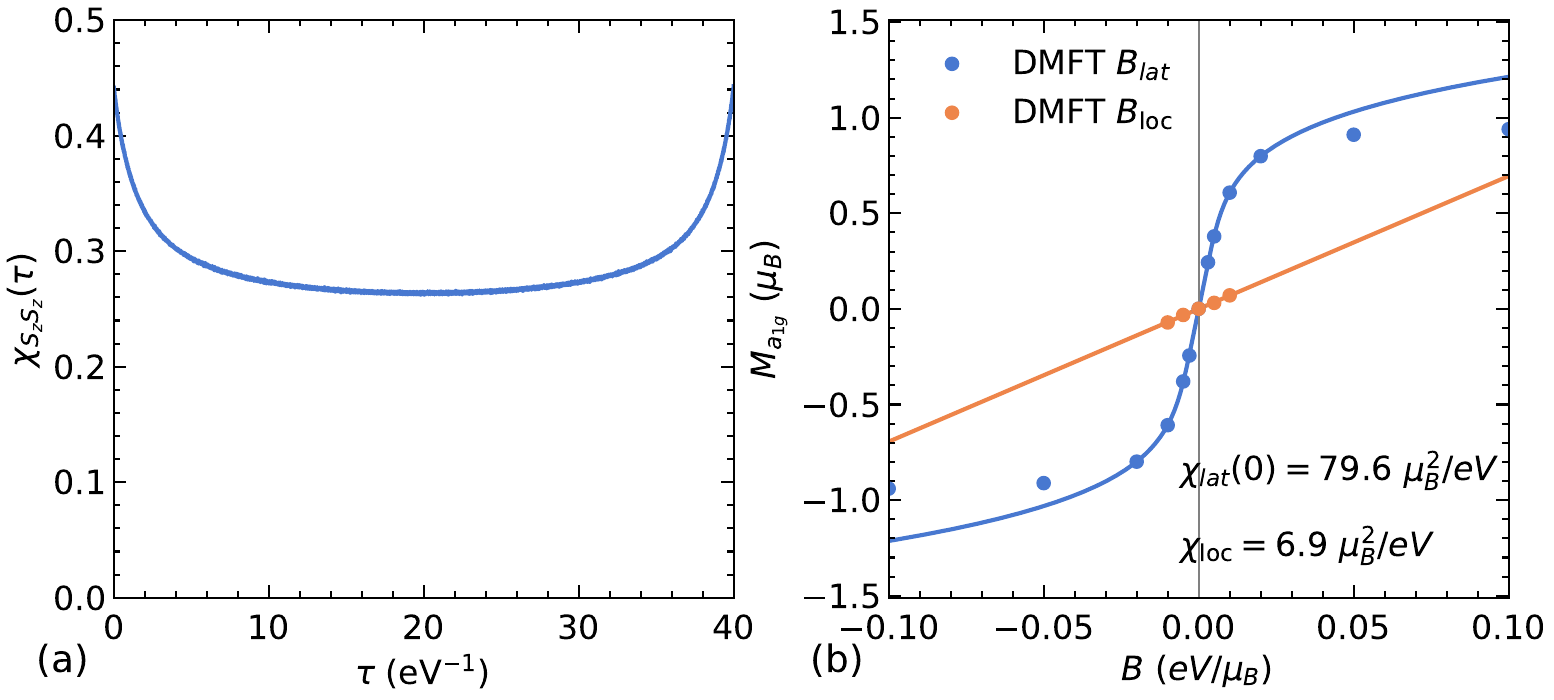}
\caption{(a) Impurity susceptibility $\chi_{S_z S_z}(\tau)$ measured at $T \!=\! 290.1$~K via QMC at the cRPA interaction values.
(b) Magnetic moment as function of an applied magnetic field locally and on the lattice calculated via QMC at $T \!=\! 290.1$~K using cRPA interaction values. Lines show a first-order ($B_\mathrm{loc}$) and fifth-order ($B_\mathrm{lat}$) polynomial fit.}
\label{sfig:susz}
\end{figure}

In the calculations shown in the main text, we focused only on the non-magnetic solution of the DMFT equations; technically this is 
done by enforcing a paramagnetic state in each DMFT iteration. We also performed calculations allowing for ferromagnetic ordering and 
observed a clear tendency of the system to order ferromagnetically (FM) just below 150~K, as shown in Fig.~\ref{sfig:magnetism}(a). 
From Fig.~\ref{sfig:magnetism}(b), it can be observed that the ordering temperature depends strongly on the Hund coupling $J$. 
We also tried to stabilize simple antiferromagnetic order by solving four independent impurity problems in the unit cell 
without any sign of ordering. 

In Fig.~\ref{sfig:susz}, we display for $T=290$~K the imaginary-time dependence of the local susceptibility (panel (a)) as well as (panel (b)) 
the local ($\chi_{imp}$) and $\mathbf{q} \!=\! 0$ uniform ($\chi_{lat}$) static susceptibilities. The latter are obtained by applying 
an external Zeeman field, either locally or uniformly, and computing the resulting magnetization. 

We observe that the local susceptibility tends to a constant at large time, indicating the 
presence of local moments at the temperature considered (which is way above the onset of coherence). 
Furthermore, the uniform susceptibility is larger than the local one, indeed consistent with a tendency to ferromagnetic inter-site correlations. 

These observations raise questions that should be addressed in future work and most likely require to go beyond a single-site 
DMFT computation. The presence of local moments above the onset temperature for coherence is indeed consistent with 
experimental observations, such as the Curie--Weiss form of the magnetic susceptibility~\cite{Urano2000} and observations from 
NMR~\cite{Shimizu2012} and $\mu$SR~\cite{Okabe_2019}. The fact that Zn substitutions on the lithium sites induce spin-glass ordering 
is also revealing in this respect~\cite{Kondo2000}. In contrast, the tendency to ferromagnetic correlations is at odds with neutrons and NMR experiments 
which reveal a dominantly antiferromagnetic exchange, and no long-range order has been reported down to the lowest temperature 
investigated (which motivates our calculations focused on the paramagnetic phase). 
We conjecture that strong spatial fluctuations in this frustrated system must be taken into account to properly describe its 
magnetic properties, hence providing motivation for future work using either cluster or vertex-based extensions of DMFT.  

\bibliography{literature}